\documentclass[preprint,3p]{elsarticle}

\usepackage{natbib}
\usepackage{xspace}
\usepackage{url}
\usepackage{balance}
\usepackage{rotating}
\usepackage{placeins} 
\usepackage{lscape}
\usepackage{multirow}
\usepackage{paralist}  
\usepackage[table]{xcolor}
\usepackage{geometry}

\usepackage{wasysym}
\usepackage{trfsigns}

\newcommand{\filledbox}{$\CIRCLE$}

\newcommand{\halfbox}{$\RIGHTcircle$}
\newcommand{\emptybox}{$\Circle$}

\newtheorem{question}{RQ}
\newtheorem{hypothesis}{Hypothesis}

\usepackage{amssymb}

\journal{Information and Software Technology}

\begin{document}

\begin{frontmatter}

\title{Field Study on Requirements Engineering:\\ Investigation of Artefacts,
Project Parameters, and Execution Strategies}

\author[label1]{Daniel M\'{e}ndez Fern\'{a}ndez\corref{cor1}}
\ead{mendezfe@in.tum.de}
\author[label4]{Stefan Wagner}
\author[label1]{Klaus Lochmann}
\author[label2]{Andrea Baumann}
\author[label3]{Holger de Carne}
 
\address[label1]{Software \& Systems Engineering, Institut f\"ur Informatik, Technische Universit\"at M\"unchen, Boltzmannstr.~3, 85748 Garching, Germany}

\address[label4]{Software Engineering Group, Institute of Software Technology, University of Stuttgart, Universit\"atsstr.~38, 70569 Stuttgart, Germany}

 \address[label2]{Fakult\"at f\"ur Elektrotechnik und Technische Informatik, Universit\"at der Bundeswehr M\"unchen, Werner-Heisenberg-Weg 39, 85577 Munich, Germany}

\address[label3]{Capgemini Technology Service Deutschland, Carl-Wery-Str. 42, 81739 Munich, Germany}

\cortext[cor1]{Corresponding author}

\begin{abstract}

\textbf{Context:} Requirements Engineering (RE) is a critical discipline 
mostly driven by uncertainty, since it is influenced by the customer domain or by the development process model used. Volatile project environments restrict the choice of methods and the decision about which artefacts to produce in RE.

\textbf{Objective:} We aim to investigate RE processes in successful project 
environments to discover characteristics and strategies that allow us to elaborate RE tailoring approaches in the future.

\textbf{Method:} We perform a field study on a set of projects at one company.
First, we investigate by content analysis which RE artefacts were produced in each project and to what extent they were produced. Second, we perform qualitative analysis of semi-structured interviews 
to discover project parameters that relate to the produced artefacts. Third, we 
use cluster analysis to infer artefact patterns and probable RE execution 
strategies, which are the responses to specific project parameters. Fourth, we 
investigate by statistical tests the effort spent in each strategy in relation to 
the effort spent in change requests to evaluate the efficiency of execution 
strategies. 

\textbf{Results:} We identified three artefact patterns and corresponding 
execution strategies. Each strategy covers different project parameters that 
impact the creation of certain artefacts. The effort analysis shows that the 
strategies have no significant differences in their effort and efficiency.

\textbf{Conclusions:} In contrast to our initial assumption that an increased effort in requirements engineering lowers the probability of change requests or project failures in general, our results show no statistically significant difference between the efficiency of the strategies. In addition, it turned out that many parameters
considered as the main causes for project failures can be successfully handled. 
Hence, practitioners can apply the artefact patterns and related project parameters 
to tailor the RE process according to individual project characteristics.

\end{abstract}


\end{frontmatter}

\section{Introduction}
\label{sec:Intro}

Requirements Engineering (RE) aims at the specification of requirements 
that reflect the purpose of a software system as well as the 
needs of all relevant stakeholders~\cite{NE00}. Since requirements are the 
critical determinants of software quality~\cite{AW05}, RE 
lays the foundation for successful development projects regarding cost and 
quality~\cite{Broy2006}. As a software engineering discipline, it contributes 
with precise and stakeholder-appropriate requirements specifications to cost-effectiveness 
in the development of a system~\cite{NE00} and, thus, RE is an 
important factor for productivity and (product) quality~\cite{DC06}.

Although a rich set of methods for RE is available, these 
methods are still not fully integrated into the development process~\citep{boehm06}. 
One reason is that RE is a broad, interdisciplinary, and 
open-ended area~\cite{Zav97}, which is driven by uncertainty. Therefore, even if 
a company defines and integrates an RE process for 
company-wide use in different projects, it still does not consider the various 
influences that engineers have to face in volatile project environments. The 
possibilities and necessities of applying available RE 
techniques are limited by different project parameters such as time, budget or 
the availability of end users. These parameters hamper a standardised and 
efficient RE process that at the same time fits individual project needs. 
The dependency on customer's capabilities and used development process models render 
the process highly variable and increases the demand to systematically 
customise the RE process according to individual project 
needs.

\subparagraph{Problem Statement.}
It is recognised that basic knowledge of RE is often missing in practice~\cite{nikula2000sps}, whereby project participants have little guidance in defining a systematic process that effectively copes with uncertain project situations. The high variability of the development processes and the diversity of RE methods makes it unclear for the practitioner which methods to apply and consequently how to design an appropriate RE process. 

To assist in the customisation of the process by applying particular methods in a certain sequence, it is fundamental to first understand the circumstances the project participants are confronted with and how they should react in the 
process according to those circumstances. Although a few first steps have been taken 
in this direction, e.g., in the area of decision-making in the context of requirements 
engineering~\cite{Aurum.2003} or in the area of activity-based customisation 
approaches~\cite{CCRGJ95, BABOK09, JE03, AW05} and situational method engineering~\cite{Brink96, AHV97}, yet missing is a fundamental understanding of
\begin{compactitem}
\item what project parameters (see Table~\ref{tab:terminology}) to consider,
\item what effects these parameters have on different execution strategies in RE (see Table~\ref{tab:terminology}) and their consequences on the quality of the produced specification documents, and
\item whether those execution strategies are efficient.
\end{compactitem}

So far, there exists little guidance on the definition of project-specific RE strategies as available studies emphasise selected aspects of RE processes, their assessment, or the general relation to project failures. Yet missing are comprehensive studies going beyond the isolated investigation of general aspects of RE processes and their maturity.

\subparagraph{Research Objective.}
To lay the foundation for an RE customisation approach that considers 
the characteristics of individual project environments, we investigate how 
RE is performed in successful projects, whether we 
can identify appropriate RE execution strategies, and how 
these strategies relate to project characteristics. 

Whereas similar project characteristics of different projects should influence the maturity of the underlying specification documents (artefacts) in a similar way, the methods used to produce those documents may vary. Thus, the variability in the process definition, i.e., the actual creation of artefacts by the use of particular methods in a particular sequence, complicates the identification, categorisation, and comparison of RE execution 
strategies that correspond to chosen project parameters. 

Therefore, we investigate RE execution strategies in a process-neutral way. For this, we analyse \emph{what} has been produced and \emph{why} it has been produced instead of exclusively focusing on \emph{how} it was produced. Having such an insight into the artefacts of single projects, we are able to objectively reproduce, categorise, and compare different RE execution strategies without having to take into account the variability of the process definitions. 

Such a process-neutral investigation allows us detailed insights into which artefacts are produced in relation to project-specific parameters and which RE execution strategies are an appropriate response to certain problems.

\subparagraph{Contribution.} 
To investigate RE execution strategies and underlying causes, we analyse real-life projects at the company Capgemini Technology Services (TS), which is specialised in custom software development projects for the application domain of business information systems. Based on this analysis, we contribute the following:
\begin{compactenum}
	\item We provide a novel, process-neutral picture of requirements engineering in real projects 
		by the degree of completeness in which requirements artefacts were produced. 
	\item We identify a set of project parameters, which influence the artefact completeness and 
		hence practical requirements engineering. This relation has not
		been analysed before.
	\item We categorise projects with artefacts of a similar degree of completeness into what we call \emph{artefact patterns}. 		This 	new analysis reflects probable execution strategies for requirements engineering
		and also how they relate to the identified project parameters.
	\item We evaluate the execution strategies by analysing the effort spent for different parts of the projects.
\end{compactenum}

A major part of the contribution is the analysis of the interrelations between these four parts. The 
discussion about which RE execution strategy should be chosen with distinctive project parameters supports practical project decisions. We finally lay a first foundation for the future elaboration of a tailoring approach, which customises RE effort according to individual project situations.

\subparagraph{Context.}

This study is performed as part of a research cooperation between the Techni\-{sche}~Universit\"at~M\"unchen and Capgemini TS, a major software and consulting company. To reach our research objective, we conduct content analyses of RE artefacts produced in 12 real-life projects, interviews with project participants, and an analysis of effort considering the effort spent in RE in relation to further effort resulting from the overall development life cycle as well as from change requests. 

\subparagraph{Terminology used in the Field Study.}
To ensure an understandable and unambiguous terminology for our paper, we explain the central terms we use to describe our study. Table~\ref{tab:terminology} summarises  terms and definitions to which we will refer in our subsequent contribution.

\begin{table}[htb]
\caption{Terminology used in this paper.
\label{tab:terminology}}
\begin{center}
\begin{tabular}{p{0.2\linewidth}p{0.7\linewidth}}
\hline
\textbf{Term} & \textbf{Description}  \\
\hline
Project & Software development effort aimed at the construction of a (software) system through the application (execution) of a development process model (see also~\cite{PS06}).\\
Development process model & Standardised organisational reference model that abstracts from the idealised execution of a development project. It includes the description of the process (definition) to follow, the work products to be generated, as well as roles involved~\cite{PS06}. Synonyms: Reference process model, methodology~\cite{ISO24744}.  \\
Process definition & Planned way of creating and modifying a set artefacts via the application of particular tasks or methods (providing structured approaches to combining different description techniques~\cite{NE00}) in a particular sequence. \\
Artefact & Deliverable that abstracts from contents of a specification document. It is used as input, output, or as an intermediate result of a process definition (see also~\cite{MPKB10}). \\
Artefact pattern & A series of sets of artefacts with similar characteristics. In this study, the sets of artefacts of different projects with a similar degree of syntactic completeness. Contrast with \emph{design pattern}~\cite{alexander77,gamma95}.\\
(RE) Execution strategy & Actual instantiation of the (RE) process definition in response to a set of project parameters, and resulting in a particular artefact pattern. \\ 
Project Parameter & Assessable condition and characteristic from inside or outside the project that influences its execution, e.g., the availability of end users. Project parameters have a direct influence on the methods used and, thus, on which artefacts to create to what extent, i.e., to which degree of completeness. \\

\hline
\end{tabular}
\end{center}
\end{table}

\subparagraph{Outline.}

The paper is organised as follows. In Section~\ref{sec:related}, we discuss related 
work in the areas of our contributions. In Section~\ref{sec:FieldStudyDesign}, we 
introduce the field study design. We give a description of the research questions, 
of the case and subject selection and of the data collection procedures before 
concluding with a brief discussion on the validity procedures. Section~\ref{sec:results} 
presents the results of the field study. We begin with a case and subject description 
and structure the results according to the formulated research questions. Finally, 
we evaluate the threats to validity. In Section~\ref{sec:conclusions}, we summarise 
our findings, describe their relation to existing evidence and their implications. We 
discuss the limitations of the study and give an outlook of the future work.

\section{Related Work}
\label{sec:related}

In the following, we discuss the work related to the four main factors we investigated
in our field study. First, we describe related work discussing and empirically analysing the degree of completeness of RE artefacts. Second, we introduce project parameters and studies identifying
such parameters. Third, we discuss execution strategies for RE and related studies on process
assessments. Finally, we describe the existing studies on the impact of RE on project efforts.

\subparagraph{Studies on Artefact Completeness.}
The completeness of requirements specifications has been discussed as an important aspect of
requirements quality~\cite{Gorschek200867}. As it is hard to assess whether a specification is complete,
in this study, we investigate the syntactic completeness of artefacts by analysing whether each artefact has been created in comparison to an abstract artefact-based reference model. We can perform this comparison with a high degree of objectivity. Kamata~and Tamai~\cite{IT07} used a similar approach and mapped existing requirements specifications to the \emph{IEEE software requirements specification Std.~830-1998}~\cite{IEEE1998}. They found  that specifications tend to be balanced in the depth they are described. To the best of our knowledge, there are no further studies performing an artefact completeness analysis.

\subparagraph{Studies on Project Parameters.}
Project parameters aim to summarise important characteristics of a project influencing its execution; for instance, the availability of end users and the expected degree of interaction between those end users and the system under consideration, both influencing the necessities of creating certain contents in the RE artefacts (e.g., use case models). 

In our study, project parameters are important aspects that we found to influence how RE was performed in a project. 
Only little work has been performed in analysing characteristics of projects in an isolated manner. The work of Aurum and Wohlin~\cite{AW05} is one example. They contribute different project parameters affecting the decisions to be taken in an RE process, e.g., stakeholder-related decisions. 

Hall, Beecham and Rainer~\cite{Hall.2002} conducted a study on RE processes with the goal of discovering 
and classifying problems. They found that 48\% of problems in the analysed software development 
projects were related to poor requirements. In addition, they discovered, although they did not use the term, a set of project parameters that caused most of the problems. Examples are developer communication, inappropriate skills, inadequate resources, staff retention, user communication, lack of training and company culture. 

Luckey et al.~\cite{luckey2010reusing} use project parameters to describes projects in a repository of
security requirements. They use these parameters to find relevant security requirements for reuse.
The found parameters range from technical parameters, such as the use of LDAP in the project, to customer
parameters, such as whether the customer is in the public sector. The study, however, does not relate
the project parameters to other factors.

\subparagraph{Studies on RE Execution Strategies.}
In general, an RE execution strategy reflects how the RE process is actually executed in a specific project. The term ``execution strategy'' arises from the area of comprehensive, customisable development process models, such as the V-Modell XT~\cite{V-Modell}, a German standard for IT development projects. The V-Modell XT offers different (project) execution strategies to be selected in a tool-supported manner when initiating a project. These strategies dictate a set of artefacts to be produced and methods to be potentially used in dependency to given project parameters, such as ``System development project (customer) with one supplier''. Development process models that include different execution strategies, however, define those strategies via coarse grained project parameters and their impact on overall artefacts and activities (e.g., considering the general creation of a requirements specification). 

To the best of our knowledge, however, an analysis of execution strategies taking into account the created RE artefacts and, in particular, their contents w.r.t. the completeness of these artefacts has not been done before. Available studies in the area of RE concentrate on the extraction of best practices and emphasise used methods and description techniques in practice (see, e.g., Boehm and Alexander~\cite{BA98},  El Enam and Madhavj~\cite{EM95} or Cox, Niazi and Verner~\cite{Cox.2009}).

\subparagraph{Studies on Effort Impacts.}
The overall goal of empirical analyses of RE in practice is to discover important impacts
of the investigated factors. Several studies analyse the impact of RE methods on 
project failures. A widely known empirical evaluation of such project failures is the \emph{Chaos Report} from the Standish Group~\cite{chaos1995} that examined project failures and related causes, such as missing user involvement. The report does not, however, give detailed insights into the study design. The \emph{Success~Study} from Buscherm\"ohle, Eckhoff and Josko~\cite{buschermohle:success} presents a similar investigation of German companies, including a description of how exactly the study was  
performed. Still, both surveys exclusively investigated failed projects and general 
causes and, thus, give no understanding on how specific problems are tackled in 
successful projects. 

Kamata~and Tamai~\cite{IT07}, who analysed artefact completeness, also investigated
the general relation to project success by determining which specific sections of the documentation 
relate to project failures. In particular, they show that the section that includes project objectives, as 
well as the section that includes the functions description, both relate to cost and time overruns in the projects. 

Similar work is presented by Sommerville and Ransom~\cite{sommerville2005esi}. 
They assessed the impact of a process improvement in RE 
at organisational level. In general, their results show ``when the RE Process 
Maturity of an organisation improved, an improvement was also observed in 
business performance indicators.'' 

Damian and Chisan~\cite{DC06} analysed process improvements in RE and the 
relation to payoffs regarding, for example, productivity and the final product quality. 
Their findings showed that improvements in RE have lead to 
(a)~improvements in developer productivity (b)~improvements in product quality (fewer 
user-reported deficiencies, fewer product defects after release) (c)~more
effective risk management.

\subparagraph{Discussion of Related Work.}

As shown in the previous sections, there exist several studies that investigate isolated aspects of RE processes, artefacts, or project parameters, and selected relations of specific factors to project failures. However, no studies are available that analyse RE in a process-neutral way and, over and above all, the relations between all the  introduced factors.

Although the benefits of process-neutral investigations are understood (see also~\cite{MPKB10}), contributions that give such insights and, thus, that would satisfy our research objectives are still missing.

\subparagraph{Previously Published Material.} 

In~\cite{MWLB10}, we performed the first step of a process-neutral investigation 
as part of a field study. We analysed the artefacts that had been produced in 
12 company projects, a set of project parameters that relate to the creation 
of the artefacts, and corresponding artefact patterns with probable requirements 
engineering execution strategies that lead to these patterns. 

In this paper, we extend the field study in two ways. First, we extend the discovered 
project parameters to a detailed taxonomy of parameters, which either enforce or 
hinder the creation of certain artefacts. Second, we evaluate the execution strategies 
with respect to the effort spent in the creation of the artefacts taking into account 
further resulting effort, e.g., spent for change requests. 
This analysis gives evidence on the efficiency of the different strategies and 
contributes to the establishment of an effective customisation approach, since 
researchers and practitioners get insights into RE in practice.

\section{Field Study Design}
\label{sec:FieldStudyDesign}

We organise the study according to Runeson and H\"ost~\cite{runeson09}: We formulate
the research questions, and describe the case and subject selection,
as well as the data collection procedures. We then define the analysis procedure, and conclude with a description of how we ensure the validity.

\subsection{Research Questions}
\label{sec:researchquestions}

The study investigates RE execution strategies and 
underlying causes reflected in a set of (re-usable) project parameters. We conduct
a process-neutral investigation, which allows us to determine and evaluate 
RE execution strategies without having to directly take 
into account the variability of the process definitions. 

In contrast to available studies, our investigation first identifies the extent 
to which the artefacts were created in different projects. We analyse 
the \emph{artefact completeness} in these projects. Based on the findings, we 
investigate which \emph{project parameters} directly influence the degree of 
completeness in the single artefacts, e.g., assessable conditions like the availability 
of end users and their effects on the creation of particular requirements 
engineering artefacts such as use case models. 

We infer probable RE execution strategies, since we 
are now able to identify \emph{artefact patterns}, i.e., similarities in the degree 
of completeness in the artefacts of different projects, resulting from potentially 
differing process definitions. We abstract from project-specific detailed 
processes used to create the artefacts, but preserve the project-specific 
characteristics as part of the project parameters. In contrast to related work 
that emphasises the analysis and assessment of RE 
processes, we identify and compare probable \emph{Requirements 
Engineering execution strategies} on the basis of the artefact patterns and 
their underlying causes, reflected in the project parameters. 

Finally, to determine whether the identified strategies were appropriate, we 
evaluate their \emph{effort impacts}: we analyse the effort spent in 
RE, change requests, and the further development life cycle. 
This gives a more detailed view on the appropriateness of the execution 
strategies than given if exclusively investigating, e.g., whether a project fails or not. 

\begin{figure}[htbp]
\begin{center}
\includegraphics[width=.5\textwidth]{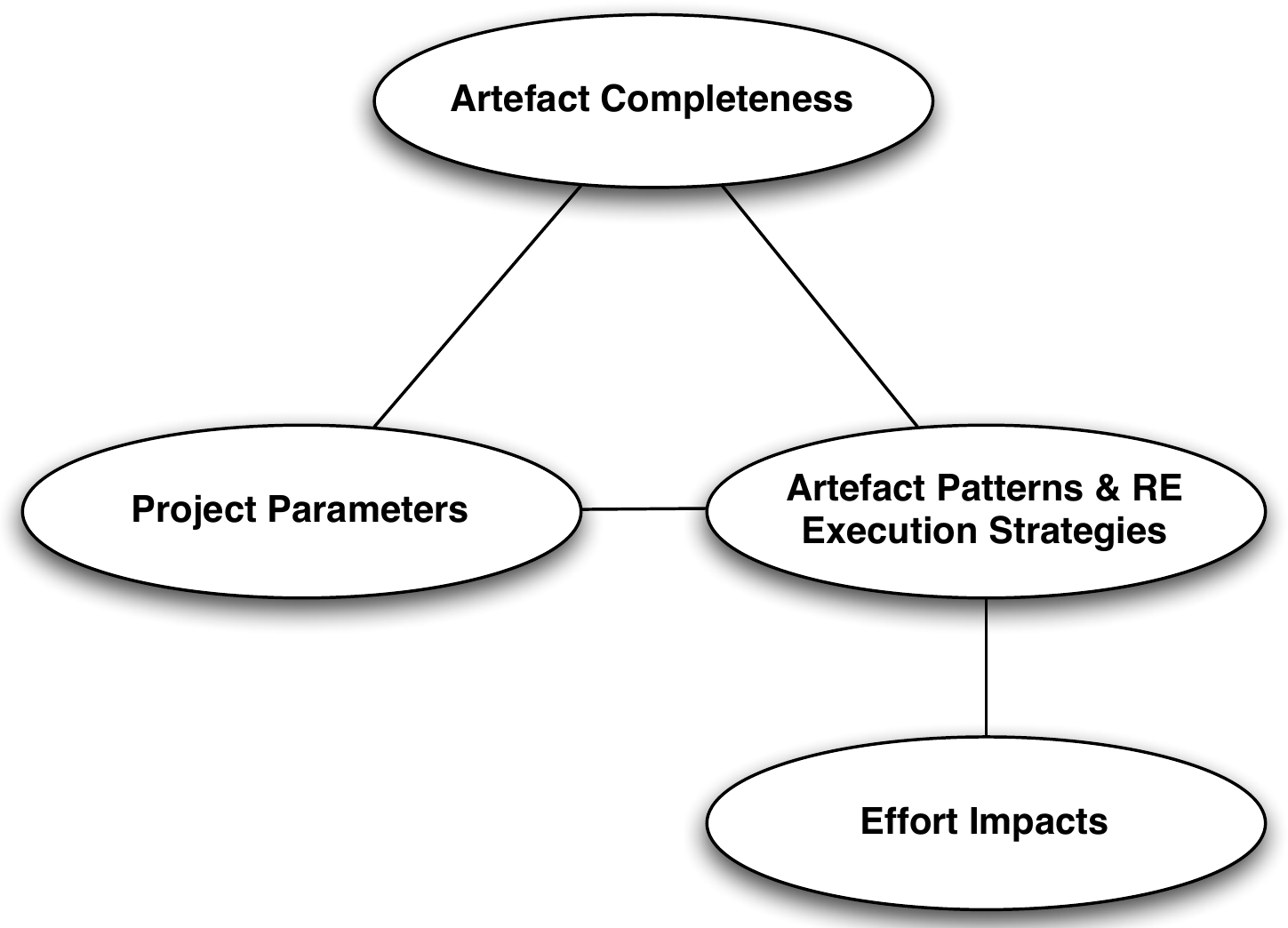}
\caption{Analysed factors and relationships.}
\label{fig:Implications}
\end{center}
\end{figure}

In summary, to reach our research objectives we have to identify and evaluate 
different execution strategies in which the artefacts are specified in a certain 
completeness in response to selected project parameters. The resulting 
factors and their relationships analysed in this field study are 
summarised in Figure~\ref{fig:Implications}.

We subsequently introduce the research questions, which we have formulated to investigate the factors.

\begin{question}
Which artefacts are created and how complete are they?\label{rq:artefact_extent}
\end{question}

As mentioned above, we need to analyse the quality of the artefacts created in different projects. This allows us to objectively reproduce, categorise, and compare RE execution strategies in different projects without having to take into account the variability of their process definitions. By abstracting from the order and way in which the artefacts were documented, we then are also able to account for the relevance of single artefacts of a particular quality with respect to specific project situations.

To determine a notion of quality in the artefacts created in a limited number of projects, we are interested in the degree of completeness in the documentation of the artefacts. With ``completeness'', we have to refer to the syntactic completeness, i.e., the elements and relations of, for example, a use case model being documented or not. The reason is that we are not able to objectively estimate the semantic completeness; for instance, whether all requirements (e.g., all use cases) relevant to the stakeholders were documented  (see also~\cite{IEEE1998}). Further information is provided in the analysis procedure in Section~\ref{sec:AnalysisProcedure}.

\begin{question}
Which project parameters have an influence on the artefact completeness?\label{rq:influences}
\end{question}

We need to find influencial project parameters and their relation to the completeness of the RE artefacts to reproducibly characterise the project-specific context of the RE process. Having identified the degree of completeness of the artefacts in individual projects, as well as influencing project parameters, we set both in relation to each other. While RQ \ref{rq:artefact_extent} addresses \emph{what} has been produced, RQ \ref{rq:influences} addresses  \emph{why} this has been produced.

\begin{question}
Are there artefact patterns and corresponding execution strategies?\label{rq:dependencies}
\end{question}

Based on the project parameters that influence individual projects in 
relation to the completeness of the RE artefacts, 
we analyse whether specific artefact patterns can be distilled. To this end, 
we compare and categorise the artefact completeness for similarities 
among different projects. We then analyse potential reasons for the artefact completeness on the basis of the identified project parameters, and infer probable RE execution strategies.

\begin{question}
Do the patterns and execution strategies differ in their effort for RE and their impact on other effort?
\label{rq:economicimpacts}
\end{question}

If we are able to identify artefact patterns with their underlying execution 
strategies, we need to evaluate their appropriateness. We are not only 
interested in whether the project is a success (in case a specific pattern is 
used with certain project parameters), but we also want to know the economic 
impact the pattern has. This allows for a comparison of the patterns and an 
identification of advantages and disadvantages of particular execution 
strategies. As a first step into this direction, we analyse whether there are 
differences between the found patterns and execution strategies with 
respect to the effort spent on RE and other effort. 
In particular, we are interested in the effort for change requests that is potentially 
caused by insufficient RE.

\subsection{Case and Subjects Selection}
\label{sec:Selection}

We analyse documented and ongoing projects\footnote{With ongoing projects we refer to projects that have further releases or increments beyond the ones analysed.} at Capgemini TS. The company constitutes the German technology service entity of the Capgemini Group and has its focus on custom software development  within the application domain of business information system. The projects to which we refer in the field study (including corresponding study subjects and objects) thus are all situated in the same application domain.

Regarding the \emph{study subjects} (the project participants), we focus on two company-specific roles, which are related to requirements engineering: ``pro\-ject lead'' and ``chief analyst''. The project lead has the responsibility of project management issues, such as control or risk management, but also of acquisition. Corresponding subjects are the source for initial information on how the development project was set up and which documents the project scope was defined upon. After the successful acquisition and scoping phase of a project, the chief analyst has the responsibility of analysing and documenting the business processes, the requirements and the initial (overall) system specifications. All the project participants that are assigned to these roles are employees of Capgemini TS. For confidentiality reasons, we have no access to customer-side project participants. 

Regarding the \emph{study objects}, we distinguish between the company-wide development process model used  at organisational level (focussing on the RE definition), and the projects, in which  the RE process is performed according to the process model. The first one serves as a preparation of the study and provides an understanding of used roles, methods, and artefacts with corresponding terminology. The study objects contain descriptions of the reference process model, training material, and standard operating procedures. With the understanding of the reference process model, we select different projects and analyse corresponding documents. As a whole, 18 projects were available and ready to participate in the study. We concentrate, however, on 12 projects (see Section~\ref{sec:CaseSubjectDescription}), since these were able to offer all data necessary to completely perform the analysis procedure (e.g., specification documents). The remaining six projects were not able to provide that data, mostly due to confidentiality issues.

The documentation of the projects includes acquisition documents, so-called study documents, requirements specifications and system specifications and, where necessary, frame contracts. Acquisition documents capture background information and possible objectives of each project. Study documents usually capture information on the customer domain and the organisation, the actual business processes, and a rough plan of the future business processes to be supported and realised by the system under consideration. These documents are mostly produced as preparation for a development project. In some cases they are created in parallel to a maintenance phase, so that they also include a detailed description of the main use cases and technical aspects of the current IT infrastructure. The requirements specification contains the user and system requirements, i.e., the external functional specification and the specification of non-functional requirements. The system specification contains the internal functional specification, which defines the logical component architecture.

Finally, the analysed projects use a standard effort accounting system that captures the effort spent by all participants for each project. The accounts within the system are organised according to the phases (respectively disciplines) of the company-specific development process model, e.g., consultancy, project management, or change management. For each of the different tasks performed in a phase, the accounts optionally include detailed sub-accounts, such as for particular quality assurance tasks. During the project, the project participants assign their effort spent to these different accounts. We use the accounting system as a source for the analysis of the effort impacts, as we do not have information about the actual costs (see also Section~\ref{sec:procedure:stat:testing} for the detailed description of the accounts and corresponding analysis procedure).

\subsection{Data Collection Procedures}
\label{sec:DataCollection}

We collect the documentation of the development process model as a preparation before analysing the single projects. The documentation was made available at the beginning of the study. We then prepare and set up the data collection from the single projects in two steps that take approximately two days for each project. This way, we keep the effort for the study subjects to a minimum. The single points of contact for the study subjects were the authors of the study, Daniel M\'{e}ndez Fern\'{a}ndez and Andrea Baumann, who also performed the interviews (see the subsequent sections).

\subparagraph{Step 1.}
As the first step, we select the projects to participate in the field study from a set of (project) candidates. To identify candidates, our project partner provides us with a list of projects and corresponding contact persons, which we contact by email. Each project is represented by one person, who either is assigned as the pro\-ject lead or as the chief analysts of the project. If the project candidate is selected, this person serves as the single point of contact during the study.

When establishing the first contact, we provide information about the study planned (its purpose and design) and propose an appointment for an open telephone interview of approximately one hour. This telephone interview is not prepared with a set of particular questions, but serves to
\begin{compactenum}
\item clarify questions of the (potential) study subjects and
\item get a brief overview of the project background, its relevance for the field study, and an overview of the existing documentation.
\end{compactenum}

We consider a project to be relevant if the project offers the possibility to access the documentation (regarding, e.g., non-disclosure agreements with third parties), the project participants are able to put in the necessary amount of effort in the study, and whether the project includes the creation of RE-relevant artefacts in the envisioned application domain of business information systems.

\subparagraph{Step 2.}

In the second step, the project participants of the selected study-relevant projects (the project lead or the chief analyst), send us the documentation for a brief content analysis. This enables an initial understanding of the artefacts (their structure and contents) and allows us to identify needs for requesting additional documents. During a second short open telephone interview, we finalise the preparation of the analysis and make an appointment for a semi-structured interview to complete the actual content analysis. 

To elaborate different project parameters that have an influence on individual 
projects and their artefacts, we perform semi-structured interviews. In each 
interview, we directly ask the study subjects for specific content items which we 
found exceptional in the content analysis. This allows a discussion on why 
specific artefacts are documented or not, and furthermore, the extraction of specific 
project parameters that are directly related to those artefacts. 

We prepare each interview by identifying those content items that exhibit differences 
compared to other projects and items that have not been specified at all. During 
this (semi-structured) interview, we then ask for:
\begin{compactenum}
\item Individual project parameters that had an effect on the particular content items
\item The dependencies that these parameters had to specific content items 
\end{compactenum}

In addition, the project participants make the information for the effort analysis available to us. This includes the project-specific names used by the project participants in the effort accounting system, since variations from the standard names are possible. Apart from the account names, they provide us with the point in time when the requirements specification was accepted by the customer. The acceptance date is used to allocate the documented effort in the effort accounting system to chosen categories, such as RE effort or effort spent in change requests (see also the next section). Based on given information, we retrieve the actual data from the accounting system and analyse it without further involvement of the project participants.

\subsection{Analysis Procedures}
\label{sec:AnalysisProcedure}

The analysis of the projects is conducted in four steps:

\begin{compactitem}
	\item[Step 1] addresses RQ~\ref{rq:artefact_extent} by content analyses.
	\item[Step 2] addresses RQ~\ref{rq:influences} by qualitative analysis of expert interviews.
	\item[Step 3] addresses RQ~\ref{rq:dependencies} by cluster analysis.
	\item[Step 4] addresses RQ~\ref{rq:economicimpacts} by statistical testing.
\end{compactitem}

The authors Daniel M\'{e}ndez Fern\'{a}ndez and Andrea Baumann conduct the first two steps and any resulting interaction with the study subjects (see also the foregoing section). Klaus Lochmann and Stefan Wagner then additionally participate in the third and fourth step. Finally, Holger de Carne performs the actual retrieval of the data from the accounting system and the removal of confidential information, e.g., details of the customers and of the project participants who assign their spent effort into the system.

In the following section, we describe each step in detail.

\subsubsection{Content Analyses (Step 1)}
This step answers RQ~\ref{rq:artefact_extent} via content analyses of the collected 
documents, which we perform in isolation without any discussions with the study subjects. Thus, we can ensure an objective review of the documentation and furthermore a neutral comparison of the documentation of different projects. In projects of a high complexity that have several thousands of requirements, we take samples for each of the documented (software) releases. This reduces the effort and gives an average view of the syntactic completeness of each artefact. Finally, requirement's attributes, such as the documented priority of requirements and details on which this priority calculation is based, are not explicitly taken into account, because they were handled differently in the projects. 

The content analysis is performed by comparing the documents and their content on the basis of a 
neutral artefact-based \emph{Requirements Engineering Reference Model} (REM)~\citep{BPKR09}. REM results from a research cooperation between the Technische Universit\"at M\"unchen and several industry partners. REM defines a taxonomy of the core RE artefacts (documents or data sets) with a description of recommended contents and an abstract description of the contents' interdependencies (traceability relations). We consider REM to be neutral, because it makes no assumptions about a particular application domain, i.e., it was not developed for a particular family of systems, nor does it relate to a particular methodology. 

In addition to the independence to a particular methodology, we opted for REM, because it also unifies in its taxonomy the RE artefacts proposed by several available artefact-based reference models, such as the \emph{IEEE software requirements specification Std.~830-1998}~\cite{IEEE1998} or the \emph{Volere requirements specification templates}~\cite{RR07}. This supports the validity of the taken reference model with respect to the state of the art in artefact orientation. Since the determination of semantically incomplete RE artefacts (e.g., if all requirements considered relevant by the stakeholders are documented) is still an unsolved issue~\cite{IEEE1998}, we focus on the syntactic completeness, i.e., the existence of the content items proposed by REM with respect to the elements and relations specified in the analysed specification documents. The investigation of the semantic completeness of the artefacts is not in scope of the field study. The only indication of semantically incomplete artefacts is given, to some extent, by the results of the analysis of the pattern efficiency (see step 4). In the analysis of the effort spent in the creation of the artefacts, we evaluate the patterns, e.g., with respect to change requests that may result from insufficient RE potentially indicating that not all requirements were documented (see also Section~\ref{sec:procedure:stat:testing}). 

We take into account the three major artefacts proposed by REM: the \emph{Business~Needs Specification},  the \emph{Requirements~Specification} and the \emph{System~Specification}. Each of those artefacts includes a list of content items. 
Taking each artefact and its included content items as a reference model, we compare the content of each projects' documents. We consider some of these content items as irrelevant for the domain of business information systems, such as hardware-specific constraints, which are relevant for embedded reactive systems. We thus tailored REM for the domain of business information systems. We performed a domain-specific interpretation of the content items and removed the irrelevant items with respect to domain-specific architecture frameworks, like the Zachmann framework (see also~\cite{Zac87, Schekkerman04}), so that no content items were missing. 

For each content item in REM, there are defined (syntactic) completeness criteria. This allows for the usage of this model as a reference for a detailed analysis of a specification's content and its completeness according to the criteria: 

\begin{compactitem}
	\item \emph{Completely specified:} The content item can be mapped unambiguously onto a specific element  of the analysed documents.
	\item \emph{Incompletely specified:} The content item can be identified in the analysed documents but exhibits major deficiencies with respect to the defined completeness criteria.
	\item \emph{Missing:} The content item cannot be identified in the analysed documents.
\end{compactitem}

The content analysis with REM gives a detailed view on the produced artefacts and their syntactic
completeness (\emph{what} has been documented). Since it allows no deeper understanding about the process and 
the underlying motivation (\emph{why} has it been documented), we perform \emph{step 2} of the analysis.

\subsubsection{Qualitative Analysis of Expert Interviews (Step 2)}

We address RQ~\ref{rq:influences} by analysing why the requirements 
are documented in a certain way. For this, we perform the interviews and use a
mixed approach to distill 
project parameters that relate to the degree of completeness in the artefacts 
(see step 2 in the data collection procedure of Section~\ref{sec:DataCollection}).
As starting theory for project parameters and categories, we employ the components 
proposed in decision (support) approaches for RE, i.e., the one from 
Aurum and Wohlin~\cite{AW05b} (see related work in Section~\ref{sec:related}).

We document each of the project parameters with its dependency to the content item it
has an impact on. We express dependencies of the parameters on content items 
as either a positive or a negative impact. Parameters can have a positive impact and thereby 
call for the creation of a content item or they can have a negative impact and hamper the 
creation of the content item. We express the first variant as ``need for action'' and the second 
variant as ``ability to act''. 

We finally remove those parameters that had different (contradictory) influences on the same 
content items over the range of examined projects. This supports the 
validity of the remaining parameters. The resulting theory is a set of project parameters in 
several categories. Structuring the project parameters according to such categories supports 
the comparison of discovered project parameters with existing studies. 

\subsubsection{Cluster Analysis (Step 3)}
After performing the content analyses in all the projects and the interviews 
with corresponding project participants, we perform a cluster analysis  to address 
RQ~\ref{rq:dependencies}. We analyse the RE artefacts of all the projects for similarities 
to identify common patterns. The analysis groups the projects into clusters with 
similar degrees of completeness in the artefacts. Hence, it is probable that the 
same RE execution strategy was performed to create the 
artefacts of projects in one group. We connect artefact patterns with 
RE execution strategies that cause the patterns. 

We perform the clustering using the k-means
cluster analysis, which minimises the distance of each project to the mean, 
called \emph{centre}, in its cluster. We encode the three possible verdicts for an 
artefact on an ordinal scale with 0 = \emph{missing}, 1 = \emph{incompletely 
specified} and 2 = \emph{completely specified}. We experiment with different 
values for the number of clusters to get a useful grouping for the relation to 
RE execution strategies. We identify suitable strategies 
by comparing the completeness of artefacts in the cluster and by analysing the 
differences between them.

\subsubsection{Statistical Testing (Step 4)}
\label{sec:procedure:stat:testing}

We perform statistical testing of the effort spent in the creation of the RE artefacts and in further development tasks to evaluate the artefact patterns identified in the foregoing cluster analysis. This step contains content analysis of the entries in the standard effort accounting system used for each project (see Section~\ref{sec:Selection}). We group the project-specific effort data from the account system into three categories:

\begin{compactitem}
	\item Effort for Requirements Engineering (REQ)
	\item Effort for change requests (CR)
	\item Effort for other tasks in the software life cycle (SWL)
\end{compactitem}

Since no explicit RE phase is assigned in the company-specific development process model, the corresponding accounts for analysing the effort data have to be interpreted (see also Section~\ref{sec:Selection}). Table~\ref{tab:accounting} summarises the accounts and how these accounts are 
allocated to the three categories used for statistical testing. 

\begin{table}[h!]
\caption{Interpretation of accounts. 
\label{tab:accounting}}
\begin{center}
\begin{tabular}{lccc}
\hline
\textbf{Accounts} & \textbf{REQ} &  \textbf{CR} & \textbf{SWL} \\
\hline
\emph{\small Specificaton} & & \\
Study current situation & B & -- & A \\
Documentation & B &  --& A  \\
Re-work & B & A & -- \\
Quality assurance & B & -- & A \\
\hline
Design & -- & -- & I \\
\hline
Realisation & -- & -- & I \\
\hline
Integration \& Test & -- & -- & I \\
\hline
Launch \& Acceptance & -- & -- & I \\
\hline
\emph{\small Project Management} & & \\
Project lead & -- & -- & I \\
Knowledge management & -- & -- & I \\
Coordination Specification \& Design & B & -- & A \\
\hline
\makebox[1.4in][l]{Infrastructure management} & -- & -- & I \\
\hline
Consultancy & B & -- & A \\
\hline
Change requests & B & A & -- \\
\hline
Cross-Cutting & -- & -- & I \\
\hline
\end{tabular}
\\[1cm]
\begin{flushleft}

\begin{tabular}{lrll}
\textbf{Legend:} & B &Before Acceptance \\ & A & After Acceptance\\ & I &  Independent of Acceptance\\

\end{tabular}

\end{flushleft}

\end{center}
\end{table}

The accounts listed in the table represent a simplification of 40~(in total) accounts and tasks. We do 
not give details of certain accounts if no differentiation between the three 
categories has to be made. For instance, the account \emph{Integration \& Test}, 
including tasks for several test phases and bugfixing, shows no details since 
all tasks can be allocated to effort spent as part of the overall software life 
cycle. The account \emph{Project Management}, on the other hand, does give details on further 
tasks, like \emph{Project Lead} or \emph{Coordination SP\&D}. The details are 
necessary, because the latter task \emph{Coordination Specification \& Design} 
is of interest. It describes the coordination of the activities in the 
specification and the design phase, in which, e.g., possible solutions for specific requirements are analysed and discussed in team meetings. This task therefore includes, in parts, effort that is to be allocated to the REQ category. 

We interpret the documented effort according to the acceptance date of the 
specification document. Hence, some accounts include effort spent for either 
RE, change requests or the general software life cycle. 
We illustrate this interpretation by using three different characters: ``B''  
refers to an interpretation of the effort as RE effort 
before acceptance of the corresponding specification; ``A'' refers to efforts
that differ in their interpretation after accepting the specification;  ``I'' 
refers to whole accounts in which we make no differentiation, such as travel 
times (belonging to the account \emph{Cross-Cutting}) that cannot be clearly 
allocated to a concrete phase. The effort documented in the account \emph{Consultancy}, 
for example, is allocated to the REQ category if it is documented before 
the specification document has been accepted by the customer. After this 
acceptance, we declare the accounted effort to belong to the software life cycle. 
Similarly, the account \emph{Change Requests} is only interpreted as regular 
change requests and thereby allocated to the corresponding category, if the requirements
specification has been accepted. 

Unfortunately, we do not have access to any consistently collected size or complexity measures, such as function points or lines of code, that could be used to normalise the effort data over all considered projects. Therefore, we define four key measures for the further analysis that are normalised differently. First, the effort data from the three categories REQ, CR, and SWL are set in relation to the total effort (TOT). This gives comparable data about what fraction of the total effort is spent on RE, change requests, and further activities. Summing up, we have the three metrics REQ/TOT, CR/TOT, and SWL/TOT. 

Second, we define a surrogate measure for the efficiency of the RE execution to be able to analyse if the investment in RE was beneficial. Ideally, RE should cover (in the end) all stakeholder needs in corresponding specifications. Later change requests are signs that the requirements engineers have failed in this and the stakeholders add additional requirements, which is usually more costly than getting it right the first time. By spending more effort on RE, we expect less change requests. Therefore, the relationship between RE effort and change request effort needs to be balanced. We can analyse this relationship by plotting CR against REQ. In addition, we define a fourth measure, CR/REQ, which expresses how much change request effort had to be spent per spent RE effort. We need this additional measure, because we are not only interested in the nature of the relationship between CR and REQ but also in the differences in this relationship over the artefact patterns.

To answer RQ~\ref{rq:economicimpacts}, we analyse the differences in the defined measures
for the artefact patterns. We are interested in whether specific RE execution strategies (reflected in the artefact patterns) tend to spend more effort on RE or change requests, and whether this had an influence on any further effort. Moreover, if one pattern tends to be more efficient in terms of our surrogate measure, it would be a candidate for further research about the reasons for this difference. If not, we have to assume that the patterns are at the same level of efficiency.

The analysis is performed by statistical testing of the differences between the distribution of the
measures. Depending on the properties of the sample data, we analyse it by parametric or nonparametric tests
at a confidence level of 0.05.
The null hypotheses we test are the following:
\begin{hypothesis}
There is no difference between the distributions of REQ/TOT in the patterns.\label{hyp:req}
\end{hypothesis}
\begin{hypothesis}
There is no difference between the distributions of CR/TOT in the patterns.\label{hyp:cr}
\end{hypothesis}
\begin{hypothesis}
There is no difference between the distributions of SWL/TOT in the patterns.\label{hyp:swl}
\end{hypothesis}
\begin{hypothesis}
There is no difference between the distributions of CR/REQ in the patterns.\label{hyp:cr_req}
\end{hypothesis}

\subsection{Validity Procedures}
\label{sec:ValidityProcedure}
The selection of projects from different industrial sectors with different sizes
and different project participants increases the external validity of the results.
Initial discussions with the study subjects lower barriers and
ensure the collection of accurate and appropriate data. The neutral artefact
model REM serves as a basis for classification, and the classifications as well
as parameter categories are reviewed by other researchers. We present and discuss the results of the overall analysis during last feedback meetings with all project participants. This clarifies open questions and excludes wrong results that could arise from incomplete documentation or wrong assumptions.

\section{Results}
\label{sec:results}

In this section, we present the results of the study. We order them by (1) the 
case and subject description, and (2) according to the results for each of the defined research questions. 

\subsection{Case and Subject Description}
\label{sec:CaseSubjectDescription}
The analysed development projects are all custom software development projects within the 
application domain of business information systems. Although they are restricted to this application domain, they exhibit different characteristics. Table \ref{tab:project_overview} gives an overview of the analysed projects. 

\begin{table}[h!]
\caption{Analysed projects (anonymised).
\label{tab:project_overview}}
\begin{center}
\begin{tabular}{p{0.28\linewidth}p{0.35\linewidth}p{0.2\linewidth}}
\hline
\textbf{ID} & \textbf{Industrial Sector}  & \textbf{Size} \\
\hline
P1  (finished) & Finance                   & Small \\
P2  (finished) & Finance                   & Small \\
P3  (finished) & Finance                   & Small \\
P4  (ongoing)  & Retail sale               & Medium \\
P5  (ongoing)  & Contracting authority          & Medium \\
P6  (ongoing)  & Telecommunication         & Large\\
P7  (ongoing)  & Logistics                 & Large \\
P8  (ongoing)  & Logistics                 & Large \\
P9  (finished) & Aerospace                 & Medium \\
P10 (ongoing)  & Contracting authority          & Medium \\
P11 (finished) & Finance                   & Medium \\
P12 (ongoing)  & Automotive                & Large \\
\hline
\end{tabular}
\end{center}
\end{table} 

For each project, we distinguish between the industrial sectors of the customers and the project size, that for reasons of confidentiality is clustered into 3 categories. We classify the projects with a size of up to 20 person years as small-scale projects, from 20 to 120 person years as medium-scale projects, and above 120 person years as large-scale projects. The analysed projects are labelled with numbers and it is also mentioned whether they are finished or still ongoing (in terms of further releases or increments). All of the projects focus on the replacement of legacy systems, on the development of new systems, or on consultancy in which application landscapes are analysed and re-designed. Consequently, all projects have in common at least the definition of requirements and system specification artefacts independently of the following phases.  Those artefacts are captured in different formats, including  MS Word, Excel and (UML) Enterprise Architect models.

Regarding the company-wide development process model and the conformance of the projects to it, nearly all projects initially followed a waterfall model as a consequence of multi-staged bidding procedures. The only exception is P4, which followed an iterative and incremental process definition right from the beginning as part of a follow-up project with no explicit bidding procedure. 

The followed development process model is a collection of architecture-driven design methods, which is based on the \emph{Integrated Architecture Framework (IAF)}~\cite{IAF}. These methods were employed by all projects as part of a so-called specification discipline according to the company-specific derivate of the development process model \emph{Rational Unified Process (RUP)}~\cite{JBR99}, but with no explicit assignment of an RE phase.

\subsection{Documented Requirements Artefacts (RQ~\ref{rq:artefact_extent})}
\label{sec:results_rq1}

For each analysed project, we compare the existence and syntactic completeness of the artefacts proposed by REM to the content of the analysed documents (see also Section~\ref{sec:AnalysisProcedure}). The effort spent for the analysis of the documents varies in dependency to the project size and ranges from approximately 4 hours for small-scale projects to 24 hours for large-scale projects. Table~\ref{tab:artefacts} on page~\pageref{tab:artefacts} summarises the results of the content analysis. 

\begin{table*}[!p]
\caption{Completeness of artefacts in the analysed projects.}
\small
\begin{tabular}{cclccccccccccccc}
%
\hline
\multicolumn{2}{l}{Project} & & P1 & P2 & P3 & P4 & P5 & P6 & P7 & P8 & P9 & P10 & P11 & P12 \\

\hline

\multirow{10}{*}{\begin{sideways}\small Business Needs Artefacts\end{sideways}} &  & Business Objectives  & \halfbox & \halfbox & \halfbox  & \filledbox & \filledbox & \filledbox & \halfbox  & \halfbox  & \filledbox & \filledbox & \filledbox & \filledbox \\
& & Customer/Market Requirements  & \filledbox & \filledbox & \filledbox & \filledbox & \filledbox & \filledbox & \filledbox & \filledbox & \filledbox & \filledbox & \halfbox  & \filledbox \\
& & Value to the Customer  & \halfbox  & \halfbox  & \halfbox  & \filledbox & \halfbox  & \emptybox & \halfbox  & \emptybox & \halfbox  & \filledbox & \filledbox & \filledbox \\
& & Main Features  & \halfbox  & \halfbox  & \halfbox  & \halfbox  & \halfbox  & \filledbox & \halfbox  & \halfbox  & \halfbox  & \halfbox  & \halfbox  & \filledbox \\
& & Assumptions Dependencies  & \halfbox  & \halfbox  & \halfbox  & \halfbox  & \filledbox & \halfbox  & \halfbox  & \halfbox  & \filledbox & \emptybox & \halfbox  & \filledbox \\
& & Scope  and Limitations & \filledbox & \filledbox & \filledbox & \emptybox & \emptybox & \halfbox  & \emptybox & \filledbox & \emptybox & \filledbox & \filledbox & \filledbox \\
& & ROI Calculation  & \halfbox  & \halfbox  & \halfbox  & \halfbox  & \filledbox & \halfbox  & \emptybox & \emptybox & \emptybox & \filledbox & \halfbox  & \halfbox  \\
& & Business Risk Analysis  & \halfbox  & \halfbox  & \halfbox  & \emptybox & \emptybox & \filledbox & \emptybox & \emptybox & \emptybox & \filledbox & \halfbox  & \halfbox  \\
& & Risk Calculation  & \filledbox & \filledbox & \filledbox & \filledbox & \halfbox  & \filledbox & \emptybox & \halfbox  & \halfbox  & \filledbox & \halfbox  & \filledbox \\
& & System Success Factors  & \emptybox & \emptybox & \emptybox & \halfbox  & \emptybox & \halfbox  & \halfbox  & \halfbox  & \halfbox  & \halfbox  & \emptybox & \halfbox  \\
\hline
\multirow{13}{*}{\begin{sideways}\small Req. Specification Artefacts\end{sideways}} & \multirow{4}{*}{\begin{sideways}\small Functional\end{sideways}} & Application Scenarios  & \halfbox  & \halfbox  & \halfbox  & \halfbox  & \filledbox & \halfbox  & \halfbox  & \filledbox & \filledbox & \halfbox  & \filledbox & \halfbox \\
& & User Interface  & \emptybox & \emptybox & \emptybox & \emptybox & \filledbox & \filledbox & \filledbox & \filledbox & \filledbox & \filledbox & \filledbox & \halfbox \\
& & User Classes  & \filledbox & \filledbox & \filledbox & \filledbox & \filledbox & \filledbox & \halfbox & \filledbox & \filledbox & \filledbox & \halfbox & \filledbox \\
& & System Interaction  & \halfbox & \halfbox & \halfbox & \emptybox & \halfbox & \filledbox & \halfbox & \filledbox & \halfbox & \halfbox & \halfbox & \halfbox \\
\cline{2-15}
 & \multirow{9}{*}{\begin{sideways}\small Non-Functional\end{sideways}} & Release Strategy  & \filledbox & \filledbox & \filledbox & \filledbox & \filledbox & \filledbox & \filledbox & \filledbox & \filledbox & \filledbox & \filledbox & \filledbox \\
& & Domain Model  & \filledbox & \filledbox & \filledbox & \filledbox & \filledbox & \filledbox & \filledbox & \filledbox & \filledbox & \filledbox & \filledbox & \filledbox \\
& & Environment Model  & \filledbox & \filledbox & \filledbox & \filledbox & \filledbox & \filledbox & \filledbox & \filledbox & \filledbox & \filledbox & \halfbox & \filledbox \\
& & System Boundaries  & \halfbox & \halfbox & \halfbox & \halfbox & \filledbox & \filledbox & \filledbox & \filledbox & \filledbox & \filledbox & \filledbox & \filledbox \\
& & Quality Requirements  & \halfbox & \halfbox & \halfbox & \emptybox & \halfbox & \halfbox & \halfbox & \halfbox & \halfbox & \halfbox & \halfbox & \halfbox \\
& & Assumptions  & \emptybox & \emptybox & \emptybox & \halfbox & \halfbox & \halfbox & \emptybox & \halfbox & \halfbox & \emptybox & \emptybox & \filledbox \\
& & SW Design Constraints  & \filledbox & \filledbox & \filledbox & \filledbox & \filledbox & \emptybox & \emptybox & \halfbox & \filledbox & \emptybox & \halfbox & \filledbox \\
& & Acceptance Criteria  & \halfbox & \halfbox & \halfbox & \emptybox & \filledbox & \halfbox & \emptybox & \emptybox & \filledbox & \emptybox & \halfbox & \emptybox \\ 
& & Acceptance Test Cases  & \emptybox & \emptybox & \emptybox & \emptybox & \filledbox & \emptybox & \emptybox & \emptybox & \filledbox & \emptybox & \emptybox & \emptybox \\
\hline
\multirow{14}{*}{\begin{sideways}\small System Specification Artefacts\end{sideways}} & \multirow{11}{*}{\begin{sideways}\small Design Concept\end{sideways}} & Release Planning  & \filledbox & \filledbox & \filledbox & \filledbox & \filledbox & \filledbox & \filledbox & \filledbox & \filledbox & \filledbox & \filledbox & \filledbox \\
& & Behaviour Model  & \filledbox & \filledbox & \filledbox & \filledbox & \filledbox & \filledbox & \filledbox & \filledbox & \filledbox & \filledbox & \filledbox & \filledbox \\
& & System Interaction  & \filledbox & \filledbox & \filledbox & \filledbox & \filledbox & \filledbox & \filledbox & \filledbox & \filledbox & \filledbox & \filledbox & \filledbox \\
& & Service Interaction  & \filledbox & \filledbox & \filledbox & \filledbox & \filledbox & \filledbox & \filledbox & \filledbox & \filledbox & \filledbox & \filledbox & \filledbox \\
& & Data Model  & \filledbox & \filledbox & \filledbox & \filledbox & \filledbox & \filledbox & \filledbox & \filledbox & \filledbox & \filledbox & \filledbox & \filledbox \\
& & User Interface  & \filledbox & \filledbox & \filledbox & \filledbox & \filledbox & \filledbox & \filledbox & \filledbox & \filledbox & \filledbox & \filledbox & \filledbox \\
& & Communication Interfaces  & \filledbox & \filledbox & \filledbox & \filledbox & \filledbox & \filledbox & \filledbox & \filledbox & \filledbox & \filledbox & \filledbox & \filledbox \\
& & Interfaces to Service Components  & \filledbox & \filledbox & \filledbox & \filledbox & \filledbox & \filledbox & \filledbox & \filledbox & \filledbox & \filledbox & \filledbox & \filledbox \\
& & Architecture Constraints  & \filledbox & \filledbox & \filledbox & \filledbox & \filledbox & \filledbox & \filledbox & \filledbox & \filledbox & \filledbox & \filledbox & \filledbox \\
& &Deployment Constraints & \filledbox & \filledbox & \filledbox & \filledbox & \filledbox & \filledbox & \filledbox & \emptybox & \filledbox & \filledbox & \filledbox & \filledbox \\
& & Coding Standards  & \filledbox & \filledbox & \filledbox & \filledbox & \filledbox & \filledbox & \filledbox & \filledbox & \filledbox & \filledbox & \filledbox & \filledbox \\
\cline{2-15}
& \multirow{3}{*}{\begin{sideways}\small Test \end{sideways}}   & Functional Test Criteria  & \filledbox & \filledbox & \filledbox & \emptybox & \filledbox & \filledbox & \halfbox & \filledbox & \filledbox & \filledbox & \halfbox & \filledbox \\
& & Integration Test Criteria  & \filledbox & \filledbox & \filledbox & \emptybox & \filledbox & \filledbox & \halfbox & \halfbox & \filledbox & \filledbox & \halfbox & \filledbox \\
& & Design Constraints Test Criteria  & \emptybox & \emptybox & \emptybox & \emptybox & \filledbox & \filledbox & \halfbox & \halfbox & \filledbox & \filledbox & \halfbox & \filledbox \\
\hline
\multicolumn{2}{l}{Traceability} & Needs~to~Requirements & \emptybox & \emptybox & \emptybox & \emptybox & \emptybox & \emptybox & \emptybox & \emptybox & \emptybox & \filledbox & \filledbox & \filledbox \\
& & Requirements~to~Sys.Spec & \emptybox & \emptybox & \emptybox & \emptybox & \emptybox & \emptybox & \filledbox & \filledbox & \filledbox & \halfbox & \halfbox & \halfbox \\
\hline
\end{tabular}
\\[1cm]
\begin{flushleft}

\begin{tabular}{lrll}
\textbf{Legend:} & {\filledbox}& Completely specified \\ & {\halfbox}& Incompletely specified\\ & {\emptybox}&  Missing\\

\end{tabular}

\end{flushleft}

\label{tab:artefacts}

\end{table*}

Going from top to bottom, the table structures the content items of REM according to the three major artefacts:  business needs, requirements specification, and system specification.In content items of the requirements specification, we distinguish between functional aspects and non-functional aspects. The first includes, for example, requirements-specific application scenarios, such as use case models. The non-functional models include, for example, the specification of quality requirements, models of the application's future environment or process constraints regarding the delivery of the application (``release strategy''). Within the system specification we discriminate between the design concepts and test case specifications. 

The bottom part of the table highlights traceability aspects, describing the maintained interdependencies between specific content items. We distinguish between traceability from the contents of the business needs to the requirements specification and from the requirements to the system specifications. We do not differentiate between forward tracing and backward tracing. In general, we observed that the content items of the initial specifications are documented in different degrees of completeness. The closer we come to the system specification, the more homogeneous and detailed they are specified. In fact, the content items of the system specifications are nearly all completely specified.

Within the business needs, nearly all projects specified the business objectives of a customer and corresponding high-level requirements, e.g., the goals (the ``Customer Requirements''). We observe, however, a highly variable handling of the further content items of the business needs. Especially the ``Return of Investment (ROI) calculations'', and the elaboration of the ``value to the customer'' was documented very differently. We observed the same for the ``Scope and Limitations'' which either were completely specified or missing. The ``System Success Factors'', including the basis for prioritising the requirements, were either incomplete or missing.

Within the requirements specification, we also found a difference in the handling of the functional analysis models, especially within the ``Application Scenarios''. Considering the non-functional analysis models, the system environment and the related content items were documented in a high degree of completeness. Instead, the rest of the content items, such as the quality requirements or the assumptions, were often incomplete. 

Finally, the system specifications are nearly all completely specified. Only the test case specifications are documented very differently.

\subsection{Influencing Project Parameters (RQ \ref{rq:influences})}
\label{sec:resultsrq2}

Within the semi-structured interviews, we built a theory about the 
project parameters with an influence on the artefact completeness. 
Table~\ref{tab:projectfactors} on page~\pageref{tab:projectfactors} summarises 
the categorised project parameters in which the interviewees participating in the 
different projects made no contradictory statements. 

The resulting table is organised according to the documented content items (see also Table~\ref{tab:artefacts}) and on the left side by the project parameter, since the identification of the project parameters was performed according to the content items (see also the analysis procedure in Section~\ref{sec:AnalysisProcedure}). We group the parameters into three major categories: The category \emph{Customers' Domain} groups parameters that arise from the industrial sector or from circumstances of a customers' organisation, such as stakeholder-specific characteristics. The category \emph{System under Consideration} groups parameters that arise from the envisioned family of systems and other characteristics of this kind. The third category \emph{Cross-Cutting Process Aspects} groups further parameters that restrict the project-specific application of the development process model. 

Each cell of the table contains the impact of one parameter on one content item as it was stated by the interviewees. The  symbol ``$+$'' indicates that the parameter has a positive impact on the content item, i.e., that the parameter can be taken as one reason for specifying the content item (``need for action''). The symbol ``$-$'' indicates that the parameter has a negative impact on a content item and thereby hampers its specification (``ability to act''). An empty cell indicates that the interviewees did not report an influence of the respective project parameter on the artefact.

We subsequently describe the parameters and their impacts. Due to the high variety of the parameters and their impacts, we organise the following description according to the three categories.

\subparagraph{Customer Domain.}
One observation is that three major circumstances arise from the customer domain: the industrial sector, the general relationship with the customer and characteristics of the different stakeholders. Especially whether project participants have weak access to business process information (e.g., for reasons of confidentiality) has an influence on the ability and the necessity to specify the ROI calculations, the value for the customer and the business risk analysis. 

In particular, many customers do not share enough details about their organisation (e.g., their business processes) for performing such calculations, mostly for reasons of confidentiality. This is especially true when elaborating the value of single requirements for the customers. Consequently, the prioritisation of requirements (expressing the relevance of requirements  for customers) can often not be performed. The only exception are projects that are performed with German governmental authorities. 
The reason is that those customers and related development projects are required to conform to the V-Modell XT~\cite{V-Modell}.  This standard demands an ROI calculation before performing a requirements analysis. 

The relationship with the customer also has an influence on the definition of the scope and the limitations, as well as the system success factors. If a customer is well-known, for example from a previous project, the limitations are left incomplete to improve efficiency. 
Similarly, the system success factors have a strong dependency on the knowledge about the customers and their domain. In particular, the less familiar customers are (e.g., in the first development project for this customer), the higher the probability of defining system success factors and also acceptance criteria. 

Further parameters that the interviewees mentioned consider the stakeholders' characteristics. The availability of the stakeholders (in particular of the future users of a system) strongly influences the possibilities of specifying application scenarios, e.g., via use cases. Projects that have no access to the users often document services (abstract description of system functions) as an abstraction of use cases with no ordered interaction scenarios. The unavailability also negatively affects the possibility of specifying content items that are directly related to application scenarios; for instance, the user classes (actors), the environment model, or the quality requirements. Consequently, this parameter implies the additional calculations of project risks.

\begin{landscape}
\begin{table}[!t]
\caption{Project parameters in the analysed projects.}
~\\[0.3cm]~
\raggedleft
\scalebox{0.7}{
\begin{tabular}{||c||p{7cm}||p{4pt}|p{4pt}|p{4pt}|p{4pt}|p{4pt}|p{4pt}|p{4pt}|p{4pt}|p{4pt}|p{4pt}||p{4pt}|p{4pt}|p{4pt}|p{4pt}||p{4pt}|p{4pt}|p{4pt}|p{4pt}|p{4pt}|p{4pt}|p{4pt}|p{4pt}|p{4pt}||p{4pt}|p{4pt}|p{4pt}|p{4pt}|p{4pt}|p{4pt}|p{4pt}|p{4pt}|p{4pt}|p{4pt}|p{4pt}||p{4pt}|p{4pt}|p{4pt}||p{4pt}|p{4pt}||}
\hline\hline
~

 &  
\multicolumn{1}{c||}{
~
}
 &  
\multicolumn{10}{c||}{
Business Needs Artefacts~
}
 &   
\multicolumn{13}{c||}{
Requirements Specification Artefacts~
}
 &  
 
\multicolumn{14}{c||}{
System Specification Artefacts~
}
 &  
 
\multicolumn{2}{c||}{
Traceability~
}
 
 \\ \cline{3-41} 
\multicolumn{1}{||c||}{
~
}
 &  
\multicolumn{1}{c||}{
~
}
 &  
\multicolumn{10}{c||}{
~
}
 &  
 
\multicolumn{4}{c||}{
Functional~
}
 &  

\multicolumn{9}{c||}{
Non-functional~
}
 &  
 
\multicolumn{11}{c||}{
Design Concept~
}
 &  
 
\multicolumn{3}{c||}{
Test~
}
 &  
 
\multicolumn{2}{c||}{
~
}
 
 \\ \hline\hline 
Category~
 &  
Project Parameter~
 &  
\rotatebox{270}{Business Objectives}
 &  
\rotatebox{270}{Customer Requirements}
 &  
\rotatebox{270}{Value to Customer}
 &  
\rotatebox{270}{Main Features}
 &  
\rotatebox{270}{Assumptions and Dependcies}
 &  
\rotatebox{270}{Scope and Limitations}
 &  
\rotatebox{270}{ROI Calculations}
 &  
\rotatebox{270}{Business Risk Analysis}
 &  
\rotatebox{270}{Risk Calculations}
 &  
\rotatebox{270}{System Success Factors}
 &  
\rotatebox{270}{Application Scenarios}
 &  
\rotatebox{270}{User Interface}
 &  
\rotatebox{270}{User Classes}
 &  
\rotatebox{270}{System Interaction}
 &  
\rotatebox{270}{Release Strategy}
 &  
\rotatebox{270}{Domain Model}
 &  
\rotatebox{270}{Environment Model}
 &  
\rotatebox{270}{System Boundaries}
 &  
\rotatebox{270}{Quality Requirements}
 &  
\rotatebox{270}{Assumptions}
 &  
\rotatebox{270}{SW Design Constraints}
 &  
\rotatebox{270}{Acceptance Criteria}
 &  
\rotatebox{270}{Acceptance Test Cases}
 &  
\rotatebox{270}{Release Planning}
 &  
\rotatebox{270}{Behaviour Model}
 &  
\rotatebox{270}{System Interaction}
 &  
\rotatebox{270}{Service Interaction}
 &  
\rotatebox{270}{Data Model}
 &  
\rotatebox{270}{User Interface}
 &  
\rotatebox{270}{Communication Interfaces}
 &  
\rotatebox{270}{Interfaces to Service Comp.}
 &  
\rotatebox{270}{Architecture Constraints}
 &  
\rotatebox{270}{Deployment Constraints}
 &  
\rotatebox{270}{Coding Standards}
 &  
\rotatebox{270}{Functional Test Criteria}
 &  
\rotatebox{270}{Integration Test Criteria}
 &  
\rotatebox{270}{Design Constr. Test Criteria}
 &  
\rotatebox{270}{Needs to Requirements}
 &  
\rotatebox{270}{Requirements to Sys.Spec}
 \\ \hline\hline 
\multirow{8}{*}{
\rotatebox{270}{Customers' Domain}
}
 &  
Governmental customer~
 &  
~
 &  
~
 &  
+~
 &  
~
 &  
~
 &  
~
 &  
+~
 &  
+~
 &  
~
 &  
~
 &  
~
 &  
~
 &  
~
 &  
~
 &  
~
 &  
~
 &  
~
 &  
~
 &  
~
 &  
~
 &  
~
 &  
+~
 &  
+~
 &  
~
 &  
~
 &  
~
 &  
~
 &  
~
 &  
~
 &  
~
 &  
~
 &  
~
 &  
~
 &  
~
 &  
~
 &  
~
 &  
~
 &  
~
 &  
~
 \\ \cline{2-41} 
\multirow{8}{*}{
\begin{sideways}~\end{sideways}
}
 &  
Weak access to business processes~
 &  
~
 &  
~
 &  
-~
 &  
~
 &  
+~
 &  
~
 &  
-~
 &  
-~
 &  
+~
 &  
~
 &  
-~
 &  
~
 &  
~
 &  
~
 &  
~
 &  
-~
 &  
~
 &  
~
 &  
~
 &  
~
 &  
~
 &  
+~
 &  
+~
 &  
~
 &  
~
 &  
~
 &  
~
 &  
-~
 &  
~
 &  
~
 &  
~
 &  
~
 &  
~
 &  
~
 &  
-~
 &  
~
 &  
~
 &  
~
 &  
~
 \\ \cline{2-41} 
\multirow{8}{*}{
\begin{sideways}~\end{sideways}
}
 &  
Weak relationship with customer ~
 &  
~
 &  
~
 &  
~
 &  
~
 &  
~
 &  
+~
 &  
~
 &  
~
 &  
+~
 &  
+~
 &  
~
 &  
~
 &  
+~
 &  
~
 &  
~
 &  
~
 &  
~
 &  
~
 &  
~
 &  
~
 &  
~
 &  
+~
 &  
+~
 &  
~
 &  
~
 &  
~
 &  
~
 &  
~
 &  
~
 &  
~
 &  
~
 &  
~
 &  
~
 &  
~
 &  
~
 &  
~
 &  
~
 &  
+~
 &  
+~
 \\ \cline{2-41} 
\multirow{8}{*}{
\begin{sideways}~\end{sideways}
}
 &  
Good relationship with customer~
 &  
~
 &  
~
 &  
~
 &  
~
 &  
~
 &  
-~
 &  
~
 &  
~
 &  
~
 &  
-~
 &  
~
 &  
~
 &  
~
 &  
~
 &  
~
 &  
~
 &  
~
 &  
~
 &  
~
 &  
~
 &  
~
 &  
~
 &  
~
 &  
~
 &  
~
 &  
~
 &  
~
 &  
~
 &  
~
 &  
~
 &  
~
 &  
~
 &  
~
 &  
~
 &  
~
 &  
~
 &  
~
 &  
~
 &  
~
 \\ \cline{2-41} 
\multirow{8}{*}{
\begin{sideways}~\end{sideways}
}
 &  
Weak knowledge of customer's domain~
 &  
~
 &  
~
 &  
~
 &  
~
 &  
~
 &  
~
 &  
~
 &  
~
 &  
~
 &  
+~
 &  
~
 &  
~
 &  
~
 &  
~
 &  
~
 &  
+~
 &  
~
 &  
~
 &  
~
 &  
~
 &  
+~
 &  
~
 &  
~
 &  
~
 &  
~
 &  
~
 &  
~
 &  
~
 &  
~
 &  
~
 &  
~
 &  
~
 &  
~
 &  
~
 &  
~
 &  
~
 &  
~
 &  
~
 &  
~
 \\ \cline{2-41} 
\multirow{8}{*}{
\begin{sideways}~\end{sideways}
}
 &  
Unavailability of stakeholders~
 &  
~
 &  
~
 &  
~
 &  
~
 &  
~
 &  
~
 &  
~
 &  
~
 &  
+~
 &  
+~
 &  
-~
 &  
-~
 &  
~
 &  
~
 &  
~
 &  
~
 &  
-~
 &  
~
 &  
-~
 &  
+~
 &  
~
 &  
+~
 &  
+~
 &  
~
 &  
~
 &  
~
 &  
~
 &  
~
 &  
~
 &  
~
 &  
~
 &  
~
 &  
~
 &  
~
 &  
~
 &  
~
 &  
~
 &  
~
 &  
~
 \\ \cline{2-41} 
\multirow{8}{*}{
\begin{sideways}~\end{sideways}
}
 &  
Weak technical ability of stakeholders ~
 &  
~
 &  
~
 &  
~
 &  
~
 &  
~
 &  
~
 &  
~
 &  
~
 &  
+~
 &  
~
 &  
~
 &  
~
 &  
~
 &  
~
 &  
~
 &  
~
 &  
~
 &  
~
 &  
-~
 &  
+~
 &  
~
 &  
~
 &  
~
 &  
~
 &  
~
 &  
~
 &  
~
 &  
~
 &  
~
 &  
~
 &  
~
 &  
~
 &  
~
 &  
~
 &  
~
 &  
~
 &  
~
 &  
~
 &  
~
 \\ \cline{2-41} 
\multirow{8}{*}{
\begin{sideways}~\end{sideways}
}
 &  
Unreliability of stakeholders ~
 &  
~
 &  
~
 &  
~
 &  
~
 &  
~
 &  
+~
 &  
~
 &  
~
 &  
+~
 &  
~
 &  
~
 &  
~
 &  
~
 &  
~
 &  
~
 &  
~
 &  
~
 &  
~
 &  
-~
 &  
~
 &  
~
 &  
+~
 &  
+~
 &  
~
 &  
~
 &  
~
 &  
~
 &  
~
 &  
~
 &  
~
 &  
~
 &  
~
 &  
~
 &  
~
 &  
~
 &  
~
 &  
~
 &  
~
 &  
+~
 \\ \hline\hline 
\multirow{10}{*}{
\rotatebox{270}{\centering\begin{minipage}{3cm}\centering System under \\Consideration\end{minipage}}
}
 &  
High degree of user interaction~
 &  
~
 &  
~
 &  
~
 &  
~
 &  
~
 &  
~
 &  
~
 &  
~
 &  
~
 &  
~
 &  
+~
 &  
+~
 &  
+~
 &  
~
 &  
~
 &  
~
 &  
+~
 &  
+~
 &  
+~
 &  
~
 &  
~
 &  
+~
 &  
~
 &  
~
 &  
~
 &  
~
 &  
~
 &  
~
 &  
+~
 &  
~
 &  
~
 &  
~
 &  
~
 &  
~
 &  
~
 &  
~
 &  
~
 &  
~
 &  
~
 \\ \cline{2-41} 
\multirow{10}{*}{
\begin{sideways}~\end{sideways}
}
 &  
High degree of innovation ~
 &  
~
 &  
~
 &  
~
 &  
~
 &  
~
 &  
+~
 &  
~
 &  
+~
 &  
~
 &  
+~
 &  
+~
 &  
~
 &  
~
 &  
+~
 &  
~
 &  
~
 &  
~
 &  
+~
 &  
+~
 &  
~
 &  
~
 &  
~
 &  
~
 &  
~
 &  
~
 &  
~
 &  
~
 &  
~
 &  
~
 &  
~
 &  
~
 &  
~
 &  
~
 &  
~
 &  
~
 &  
~
 &  
~
 &  
~
 &  
~
 \\ \cline{2-41} 
\multirow{10}{*}{
\begin{sideways}~\end{sideways}
}
 &  
Emphasis on data flow ~
 &  
~
 &  
~
 &  
~
 &  
~
 &  
~
 &  
~
 &  
~
 &  
~
 &  
~
 &  
~
 &  
-~
 &  
-~
 &  
-~
 &  
-~
 &  
~
 &  
~
 &  
~
 &  
~
 &  
~
 &  
~
 &  
~
 &  
~
 &  
~
 &  
~
 &  
~
 &  
~
 &  
+~
 &  
+~
 &  
-~
 &  
~
 &  
~
 &  
~
 &  
~
 &  
~
 &  
~
 &  
~
 &  
~
 &  
~
 &  
~
 \\ \cline{2-41} 
\multirow{10}{*}{
\begin{sideways}~\end{sideways}
}
 &  
Emphasis on control flow ~
 &  
~
 &  
~
 &  
~
 &  
~
 &  
~
 &  
~
 &  
~
 &  
~
 &  
~
 &  
~
 &  
+~
 &  
+~
 &  
+~
 &  
+~
 &  
~
 &  
~
 &  
~
 &  
~
 &  
~
 &  
~
 &  
~
 &  
~
 &  
~
 &  
~
 &  
~
 &  
+~
 &  
+~
 &  
~
 &  
+~
 &  
+~
 &  
~
 &  
~
 &  
~
 &  
~
 &  
~
 &  
~
 &  
~
 &  
~
 &  
~
 \\ \cline{2-41} 
\multirow{10}{*}{
\begin{sideways}~\end{sideways}
}
 &  
High degree of distribution~
 &  
~
 &  
~
 &  
~
 &  
~
 &  
~
 &  
~
 &  
~
 &  
~
 &  
+~
 &  
+~
 &  
+~
 &  
~
 &  
~
 &  
+~
 &  
~
 &  
~
 &  
+~
 &  
+~
 &  
-~
 &  
~
 &  
~
 &  
~
 &  
~
 &  
~
 &  
~
 &  
~
 &  
~
 &  
~
 &  
~
 &  
~
 &  
~
 &  
+~
 &  
~
 &  
~
 &  
~
 &  
~
 &  
~
 &  
~
 &  
~
 \\ \cline{2-41} 
\multirow{10}{*}{
\begin{sideways}~\end{sideways}
}
 &  
Complex dependencies on external systems~
 &  
~
 &  
~
 &  
~
 &  
~
 &  
~
 &  
~
 &  
~
 &  
~
 &  
~
 &  
~
 &  
+~
 &  
~
 &  
~
 &  
~
 &  
~
 &  
~
 &  
~
 &  
~
 &  
-~
 &  
~
 &  
~
 &  
~
 &  
~
 &  
~
 &  
~
 &  
~
 &  
~
 &  
~
 &  
~
 &  
~
 &  
~
 &  
~
 &  
+~
 &  
~
 &  
~
 &  
+~
 &  
~
 &  
~
 &  
~
 \\ \cline{2-41} 
\multirow{10}{*}{
\begin{sideways}~\end{sideways}
}
 &  
Weak knowledge of operative environment~
 &  
~
 &  
~
 &  
~
 &  
~
 &  
~
 &  
~
 &  
~
 &  
~
 &  
+~
 &  
~
 &  
~
 &  
~
 &  
~
 &  
~
 &  
~
 &  
~
 &  
-~
 &  
-~
 &  
-~
 &  
~
 &  
~
 &  
~
 &  
~
 &  
~
 &  
~
 &  
-~
 &  
~
 &  
~
 &  
~
 &  
~
 &  
~
 &  
~
 &  
~
 &  
~
 &  
~
 &  
-~
 &  
~
 &  
~
 &  
~
 \\ \cline{2-41} 
\multirow{10}{*}{
\begin{sideways}~\end{sideways}
}
 &  
Weak knowledge of op. background ~
 &  
~
 &  
~
 &  
~
 &  
~
 &  
~
 &  
~
 &  
~
 &  
~
 &  
+~
 &  
~
 &  
~
 &  
-~
 &  
~
 &  
~
 &  
~
 &  
~
 &  
~
 &  
~
 &  
-~
 &  
~
 &  
~
 &  
~
 &  
~
 &  
~
 &  
~
 &  
~
 &  
~
 &  
~
 &  
~
 &  
~
 &  
~
 &  
~
 &  
~
 &  
~
 &  
~
 &  
~
 &  
~
 &  
~
 &  
~
 \\ \cline{2-41} 
\multirow{10}{*}{
\begin{sideways}~\end{sideways}
}
 &  
Custom software~
 &  
~
 &  
~
 &  
~
 &  
~
 &  
~
 &  
~
 &  
~
 &  
~
 &  
~
 &  
~
 &  
+~
 &  
~
 &  
~
 &  
~
 &  
~
 &  
~
 &  
~
 &  
~
 &  
~
 &  
~
 &  
~
 &  
~
 &  
~
 &  
~
 &  
~
 &  
~
 &  
~
 &  
~
 &  
~
 &  
~
 &  
~
 &  
~
 &  
~
 &  
~
 &  
~
 &  
+~
 &  
~
 &  
~
 &  
~
 \\ \cline{2-41} 
\multirow{10}{*}{
\begin{sideways}~\end{sideways}
}
 &  
Standard software~
 &  
~
 &  
~
 &  
~
 &  
~
 &  
~
 &  
~
 &  
~
 &  
~
 &  
~
 &  
~
 &  
-~
 &  
~
 &  
~
 &  
~
 &  
~
 &  
~
 &  
~
 &  
~
 &  
-~
 &  
~
 &  
~
 &  
~
 &  
~
 &  
~
 &  
~
 &  
~
 &  
~
 &  
~
 &  
~
 &  
~
 &  
~
 &  
~
 &  
~
 &  
~
 &  
~
 &  
~
 &  
~
 &  
~
 &  
~
 \\ \hline\hline 
\multirow{13}{*}{
\rotatebox{270}{Cross-Cutting Process Aspects}
}
 &  
Time-boxing~
 &  
~
 &  
~
 &  
~
 &  
~
 &  
~
 &  
~
 &  
~
 &  
~
 &  
+~
 &  
~
 &  
~
 &  
~
 &  
~
 &  
~
 &  
~
 &  
~
 &  
~
 &  
~
 &  
~
 &  
~
 &  
~
 &  
~
 &  
~
 &  
+~
 &  
~
 &  
~
 &  
~
 &  
~
 &  
~
 &  
~
 &  
~
 &  
~
 &  
~
 &  
~
 &  
~
 &  
~
 &  
~
 &  
+~
 &  
+~
 \\ \cline{2-41} 
\multirow{13}{*}{
\begin{sideways}~\end{sideways}
}
 &  
Existence external parties~
 &  
~
 &  
~
 &  
~
 &  
~
 &  
~
 &  
+~
 &  
~
 &  
~
 &  
+~
 &  
~
 &  
~
 &  
~
 &  
~
 &  
~
 &  
~
 &  
~
 &  
~
 &  
+~
 &  
~
 &  
~
 &  
~
 &  
~
 &  
~
 &  
~
 &  
~
 &  
~
 &  
~
 &  
~
 &  
~
 &  
+~
 &  
~
 &  
~
 &  
+~
 &  
~
 &  
~
 &  
+~
 &  
~
 &  
~
 &  
~
 \\ \cline{2-41} 
\multirow{13}{*}{
\begin{sideways}~\end{sideways}
}
 &  
External acceptance tests~
 &  
~
 &  
~
 &  
~
 &  
~
 &  
~
 &  
~
 &  
~
 &  
~
 &  
+~
 &  
~
 &  
+~
 &  
~
 &  
~
 &  
~
 &  
~
 &  
~
 &  
~
 &  
~
 &  
+~
 &  
~
 &  
~
 &  
+~
 &  
+~
 &  
~
 &  
~
 &  
~
 &  
~
 &  
~
 &  
~
 &  
~
 &  
~
 &  
~
 &  
~
 &  
~
 &  
+~
 &  
~
 &  
~
 &  
~
 &  
+~
 \\ \cline{2-41} 
\multirow{13}{*}{
\begin{sideways}~\end{sideways}
}
 &  
Explicit assignment of RE~
 &  
~
 &  
~
 &  
~
 &  
~
 &  
~
 &  
~
 &  
~
 &  
~
 &  
~
 &  
~
 &  
~
 &  
~
 &  
~
 &  
~
 &  
~
 &  
~
 &  
~
 &  
~
 &  
~
 &  
~
 &  
~
 &  
~
 &  
~
 &  
~
 &  
~
 &  
~
 &  
~
 &  
+~
 &  
~
 &  
~
 &  
~
 &  
~
 &  
~
 &  
~
 &  
~
 &  
~
 &  
~
 &  
~
 &  
~
 \\ \cline{2-41} 
\multirow{13}{*}{
\begin{sideways}~\end{sideways}
}
 &  
High amount of requirements~
 &  
~
 &  
~
 &  
~
 &  
~
 &  
~
 &  
~
 &  
~
 &  
~
 &  
+~
 &  
~
 &  
~
 &  
~
 &  
~
 &  
~
 &  
~
 &  
~
 &  
~
 &  
~
 &  
~
 &  
~
 &  
~
 &  
~
 &  
~
 &  
~
 &  
~
 &  
~
 &  
~
 &  
~
 &  
~
 &  
~
 &  
~
 &  
~
 &  
~
 &  
~
 &  
~
 &  
~
 &  
~
 &  
+~
 &  
+~
 \\ \cline{2-41} 
\multirow{13}{*}{
\begin{sideways}~\end{sideways}
}
 &  
Long project duration~
 &  
~
 &  
~
 &  
~
 &  
~
 &  
~
 &  
+~
 &  
~
 &  
~
 &  
~
 &  
~
 &  
~
 &  
~
 &  
~
 &  
~
 &  
~
 &  
~
 &  
~
 &  
~
 &  
~
 &  
~
 &  
~
 &  
~
 &  
~
 &  
~
 &  
~
 &  
~
 &  
~
 &  
+~
 &  
~
 &  
~
 &  
~
 &  
~
 &  
~
 &  
~
 &  
~
 &  
~
 &  
~
 &  
+~
 &  
+~
 \\ \cline{2-41} 
\multirow{13}{*}{
\begin{sideways}~\end{sideways}
}
 &  
Estimations of functional complexity~
 &  
~
 &  
~
 &  
~
 &  
~
 &  
~
 &  
~
 &  
~
 &  
~
 &  
~
 &  
~
 &  
+~
 &  
~
 &  
~
 &  
~
 &  
~
 &  
~
 &  
~
 &  
~
 &  
~
 &  
~
 &  
~
 &  
~
 &  
~
 &  
~
 &  
~
 &  
~
 &  
~
 &  
~
 &  
~
 &  
~
 &  
~
 &  
~
 &  
~
 &  
~
 &  
~
 &  
~
 &  
~
 &  
~
 &  
~
 \\ \cline{2-41} 
\multirow{13}{*}{
\begin{sideways}~\end{sideways}
}
 &  
Weak given documentation~
 &  
~
 &  
~
 &  
~
 &  
~
 &  
~
 &  
~
 &  
~
 &  
~
 &  
+~
 &  
~
 &  
~
 &  
~
 &  
~
 &  
~
 &  
~
 &  
~
 &  
~
 &  
~
 &  
~
 &  
~
 &  
~
 &  
~
 &  
~
 &  
~
 &  
~
 &  
~
 &  
~
 &  
~
 &  
~
 &  
~
 &  
~
 &  
~
 &  
~
 &  
~
 &  
~
 &  
~
 &  
~
 &  
~
 &  
~
 \\ \cline{2-41} 
\multirow{13}{*}{
\begin{sideways}~\end{sideways}
}
 &  
Change mgmt. established~
 &  
~
 &  
~
 &  
~
 &  
~
 &  
~
 &  
+~
 &  
~
 &  
~
 &  
+~
 &  
~
 &  
+~
 &  
~
 &  
~
 &  
~
 &  
~
 &  
~
 &  
~
 &  
~
 &  
~
 &  
~
 &  
~
 &  
+~
 &  
~
 &  
~
 &  
~
 &  
~
 &  
~
 &  
~
 &  
~
 &  
~
 &  
~
 &  
~
 &  
~
 &  
~
 &  
~
 &  
~
 &  
~
 &  
+~
 &  
+~
 \\ \cline{2-41} 
\multirow{13}{*}{
\begin{sideways}~\end{sideways}
}
 &  
Standardised design process~
 &  
~
 &  
~
 &  
~
 &  
~
 &  
~
 &  
~
 &  
~
 &  
~
 &  
~
 &  
~
 &  
~
 &  
~
 &  
~
 &  
~
 &  
~
 &  
~
 &  
~
 &  
~
 &  
~
 &  
~
 &  
+~
 &  
~
 &  
~
 &  
+~
 &  
+~
 &  
+~
 &  
+~
 &  
+~
 &  
+~
 &  
+~
 &  
+~
 &  
+~
 &  
+~
 &  
+~
 &  
~
 &  
~
 &  
~
 &  
~
 &  
~
 \\ \cline{2-41} 
\multirow{13}{*}{
\begin{sideways}~\end{sideways}
}
 &  
Large team-size~
 &  
~
 &  
~
 &  
~
 &  
~
 &  
~
 &  
+~
 &  
~
 &  
~
 &  
~
 &  
~
 &  
+~
 &  
~
 &  
~
 &  
~
 &  
~
 &  
~
 &  
~
 &  
~
 &  
~
 &  
~
 &  
~
 &  
~
 &  
~
 &  
~
 &  
~
 &  
~
 &  
~
 &  
+~
 &  
~
 &  
~
 &  
~
 &  
~
 &  
~
 &  
~
 &  
~
 &  
~
 &  
~
 &  
~
 &  
~
 \\ \cline{2-41} 
\multirow{13}{*}{
\begin{sideways}~\end{sideways}
}
 &  
High team distribution~
 &  
~
 &  
~
 &  
~
 &  
~
 &  
~
 &  
~
 &  
~
 &  
~
 &  
+~
 &  
~
 &  
+~
 &  
~
 &  
~
 &  
~
 &  
~
 &  
~
 &  
~
 &  
~
 &  
~
 &  
~
 &  
~
 &  
+~
 &  
~
 &  
~
 &  
~
 &  
~
 &  
~
 &  
~
 &  
~
 &  
~
 &  
~
 &  
~
 &  
~
 &  
~
 &  
~
 &  
~
 &  
~
 &  
~
 &  
+~
 \\ \cline{2-41} 
\multirow{13}{*}{
\begin{sideways}~\end{sideways}
}
 &  
Weakly experiences team~
 &  
~
 &  
~
 &  
~
 &  
~
 &  
~
 &  
~
 &  
~
 &  
~
 &  
+~
 &  
~
 &  
~
 &  
~
 &  
~
 &  
~
 &  
~
 &  
~
 &  
~
 &  
~
 &  
-~
 &  
~
 &  
~
 &  
~
 &  
~
 &  
~
 &  
~
 &  
~
 &  
~
 &  
~
 &  
~
 &  
~
 &  
~
 &  
~
 &  
~
 &  
~
 &  
~
 &  
~
 &  
~
 &  
~
 &  
~
 \\ \hline\hline 
\end{tabular}
}
~\\[0.2cm]~
\raggedleft

\begin{flushleft}
\begin{tabular}{lrl}
\textbf{Legend:} &+&= Need for Action\\&-&= Ability to Act
\end{tabular}
\end{flushleft}

\label{tab:projectfactors}
\end{table} 
\end{landscape}

Similarly, the technical knowledge of the stakeholders negatively impacts the quality requirements. When specifying quality requirements, reference models and reference values are often missing. Quality requirements are thus left on a high level of abstraction. For example, instead of stating specific security requirements, customers often refer to security standards, such as the \emph{Common Criteria}. This project parameter also has a strong relation to the assumptions that then often have to be made explicit. Missing or incomplete quality requirements increase the probability of having to explicitly specify assumptions to compensate for upcoming risks relating to the quality requirements. Finally, a similar effect can be observed with regards to the reliability of stakeholders. The more unreliable the stakeholders and, thus, the higher the probability of having moving targets, the higher the probability of specifying weak quality requirements. In addition, this parameter argues for the specification of acceptance criteria.

\subparagraph{System under Consideration.}
Regarding the system under consideration, the expected degree of user interaction has a positive influence on the use of application scenarios and related content items, such as the data model (in terms of modelling objects being processed as part of an interaction scenario). Also, a high degree of innovation has a positive effect on the use of application scenarios, since these are documented if the feasibility of the system is unclear. 

Another parameter with similar effects is the type of system. Building a workflow management system, for example, has a positive effect on application scenarios. Instead, if considering, e.g., database integration or content management systems with low user interaction, the projects emphasise the specification of the data model rather than application scenarios. The degree of distribution and the degree of dependencies on surrounding systems also has a positive impact on application scenarios, but hampers the specification of quality requirements (in early stages of development).

A further negative impact on quality requirements is given by insufficient knowledge 
of the operative background of the systems. For instance, if the requirements 
engineers cannot define how many users will access the system simultaneously, 
the corresponding quality requirements will remain at a high level of abstraction. 

Finally, projects that target standard software, such as SAP components, differ strongly from ones that target custom software. This parameter influences, similar to the availability of stakeholders, the existence of application scenarios. If the project considers standard software instead of custom software, services are specified instead of detailed application scenarios.

\subparagraph{Cross-Cutting Process Aspects.}
Regarding the characteristics that arise from the development process, one remarkable influence is time-boxing, which means the project team faces a hard deadline for the milestones (maybe at the cost of quality). Projects that have agreed on time-boxing have a detailed release plan and also specify the linkage between the content items  to ensure traceability. A traceability matrix supports the control of the actual degree of realisation of the requirements, especially necessary for time-boxing, but also gives an overview of the number of requirements in general. Time-boxing is seen in general as a risk within projects and hence implies detailed risk calculations. Further risks are implied by the involvement of external parties. If acceptance tests are performed by external parties, projects document their scope and limitations, as done in acceptance criteria and functional test criteria.

Projects that have a long duration comprehensively specify a data model since 
they use data models as an ontology within a project, defining the relevant terms 
within the envisioned domain. Similar impacts on the data model are the team size, 
the degree of distribution and the experience of the team. Besides also impacting the risk calculations, the latter has a negative impact on the quality requirements, since team members are not aware of the technical consequences of quality requirements.

When performing estimations of functional complexity and effort by the use of, for example, function points, project participants consider application scenarios to be necessary. The application scenarios also provide support in change-intensive projects with unstable requirements. 

In general, application scenarios are explicitly documented within a requirements specification if either a requirements phase with an acceptance phase of corresponding specifications is assigned (instead of a continuous development project), or if a change management process is set up. In the latter case, the customers are often the ones to state functional requirements by the means of use cases. 

Finally, the completeness of the content items of the design concept has its rationale in the standardised design process of the company according to the IAF and RUP.

\subsection{Artefact Patterns (RQ \ref{rq:dependencies})}
\label{sec:resultsrq3}

Regarding the artefact patterns, we identify clusters with similar degrees of completeness in the artefacts. The k-means cluster analysis gives the most useful grouping into three artefact patterns. The Eucledian distances from the projects to their cluster centres range from 1 to 3.3 based on the encoding of completeness from~0 to~2. A higher number of clusters results in very small clusters that cannot be reasonably interpreted, but have similar distances. Table~\ref{tab:artefact_patterns} shows the mapping of projects to the artefact patterns, as well as the cluster centres from the k-means cluster analysis for those artefacts that have differing values amongst these patterns. The resulting values of the cluster centres are mapped to the range used in RQ~1 (see Table~\ref{tab:artefacts} on page~\pageref{tab:artefacts}) to express the artefact completeness: \emph{completely specifed}, \emph{incompletely specified}, and \emph{missing}.

\newlength{\tlentab}
\setlength{\tlentab}{7cm}

\begin{table}[!h]
\caption{Artefact patterns and corresponding artefacts, for which the completeness
differs between patterns.}
\centering
\begin{tabular}{l*{2}{c}*{2}{c}*{2}{c}}
\hline
Pattern & \begin{sideways}\textbf{Solution-Oriented}\end{sideways} &  \begin{sideways}P1--P4\end{sideways} & \begin{sideways}\textbf{Functional-Oriented}\end{sideways} &  \begin{sideways}P5, P9\end{sideways} & \begin{sideways}\textbf{Problem-Oriented}\end{sideways} &  \begin{sideways}P6--P8, P10--P12\end{sideways} \\ 
\hline
\makebox[\tlentab][l]{Business  Objectives} & \multicolumn{2}{c}{\halfbox} & \multicolumn{2}{c}{\filledbox} & \multicolumn{2}{c}{\filledbox}  \\
\makebox[\tlentab][l]{Assumptions} & \multicolumn{2}{c}{\halfbox} & \multicolumn{2}{c}{\filledbox} & \multicolumn{2}{c}{\halfbox}   \\ 
\makebox[\tlentab][l]{Scope} & \multicolumn{2}{c}{\filledbox} & \multicolumn{2}{c}{\emptybox} & \multicolumn{2}{c}{\filledbox}  \\ 
\makebox[\tlentab][l]{Business Risk} & \multicolumn{2}{c}{\halfbox} & \multicolumn{2}{c}{\emptybox} & \multicolumn{2}{c}{\halfbox}  \\ 
\makebox[\tlentab][l]{Risk Calculation} & \multicolumn{2}{c}{\filledbox} & \multicolumn{2}{c}{\halfbox} & \multicolumn{2}{c}{\halfbox}  \\
\makebox[\tlentab][l]{System Success Factors} & \multicolumn{2}{c}{\emptybox} & \multicolumn{2}{c}{\halfbox} & \multicolumn{2}{c}{\halfbox}  \\
\makebox[\tlentab][l]{Application Scenarios} & \multicolumn{2}{c}{\halfbox} & \multicolumn{2}{c}{\filledbox} & \multicolumn{2}{c}{\halfbox}  \\
\makebox[\tlentab][l]{User Interface} & \multicolumn{2}{c}{\emptybox} & \multicolumn{2}{c}{\filledbox} & \multicolumn{2}{c}{\filledbox}  \\
\makebox[\tlentab][l]{System Boundaries} & \multicolumn{2}{c}{\halfbox} & \multicolumn{2}{c}{\filledbox} & \multicolumn{2}{c}{\filledbox}  \\
\makebox[\tlentab][l]{Assumptions} & \multicolumn{2}{c}{\emptybox} & \multicolumn{2}{c}{\halfbox} & \multicolumn{2}{c}{\halfbox}  \\
\makebox[\tlentab][l]{SW Design Constraints} & \multicolumn{2}{c}{\filledbox} & \multicolumn{2}{c}{\filledbox} & \multicolumn{2}{c}{\halfbox}  \\
\makebox[\tlentab][l]{Design Constraints~Test Criteria}& \multicolumn{2}{c}{\emptybox} & \multicolumn{2}{c}{\filledbox} & \multicolumn{2}{c}{\filledbox}  \\
\makebox[\tlentab][l]{Acceptance Criteria} & \multicolumn{2}{c}{\halfbox} & \multicolumn{2}{c}{\filledbox} & \multicolumn{2}{c}{\emptybox}  \\
\makebox[\tlentab][l]{Acceptance Test Cases} & \multicolumn{2}{c}{\emptybox} & \multicolumn{2}{c}{\filledbox} & \multicolumn{2}{c}{\emptybox}  \\
\makebox[\tlentab][l]{Tracing: Business Needs~to~Requirements} & \multicolumn{2}{c}{\emptybox} & \multicolumn{2}{c}{\emptybox} & \multicolumn{2}{c}{\halfbox}  \\ 
\makebox[\tlentab][l]{Tracing: Requirements~to~System Specification} & \multicolumn{2}{c}{\emptybox} & \multicolumn{2}{c}{\halfbox} & \multicolumn{2}{c}{\halfbox} \\ 
 \hline
\end{tabular}
\label{tab:artefact_patterns}

\end{table}

The analysis of the three artefact patterns shows different tendencies in the artefacts. One cluster has more emphasis on the solution description, one on the functional description, and one on the problem description. This leads us to the assumption that the artefact patterns are caused by a \emph{solution-oriented}, a \emph{functional-oriented}, and a \emph{problem-oriented} RE execution strategy.

\subparagraph{Solution Orientation.}
This pattern reflects a solution-biased approach and implies a weak description of the content of the business specification. The corresponding project parameters show that this pattern results from circumstances that mostly have to do with the customers' domain, such as the \emph{relationship with customers} and the \emph{knowledge of the customer's domain}. The consequences are, for example, that the business objectives and the future system environment are incompletely specified and that the system success factors are missing. Also, the projects emphasise risk calculations and the initial scope. Another consequence is that the traceability is missing due to the incompleteness of the content that shoud be traced.

\subparagraph{Functional Orientation.}	 
The projects within this pattern emphasise the functional analysis models of the requirements specification, including the application scenarios and the user interfaces. This focus is also reflected in the traceability from the requirements to the system specification. This mainly arises from the project parameter that considers the establishment of a \emph{change management} and whether an \emph{RE phase is explicitly assigned} with direct participation of the customer. A further parameter is the \emph{availability of the stakeholders}. Corresponding projects have set up a change management process and end users have been able to contribute to the definition of functional analysis models (like use cases). Since the functional demands requested by the customers can be linked up with the elements of the system specification, this positively affects the traceability and the acceptance criteria. In contrast to the solution-oriented approach, the business objectives are completely specified.

\subparagraph{Problem Orientation.}
The projects in this pattern have, in particular, a profound specification of the business needs. The strong focus on business needs is also reflected in the traceability that includes, in contrast to the functional-oriented pattern, a linkage between the requirements and the business needs. Some major reasons for this strategy are project parameters like a \emph{high degree of innovation}, the \emph{relation to customers} enabling insights into customers' organisations and business processes, or whether it was a \emph{governmental customer}.

\subparagraph{Summary.} In summary, half of the projects act in a problem-oriented way, while a third act in a solution-oriented way, and a sixth in a functional-oriented way. Even if solution orientation is an established phenomenon in practice~\citep{boehm06}, we can depict the strategy with artefact patterns and its relation to project parameters. For instance, we observe that project parameters resulting from the customer domain have a strong influence on the choice of a solution-oriented approach. The reason is that the parameters reflect the possibilities of accessing business knowledge, and whether customers can contribute to a clear requirements analysis. If customers do not give detailed insights into their organisation and their business processes, it seems that solution orientation is a way to successfully tackle this problem. This is then reflected in the low degree of completeness in the content items within the business needs specification. At the same time, one can observe  an increasing degree of completeness in risk calculations and in the initial scope definitions. 

None of the interviewees showed awareness of having made an explicit decision about whether to follow solution orientation or not during the last feedback meeting. This unawareness of possibilities and necessities in RE may come from the company-specific development process model and/or underlying solution-biased architecture frameworks. Both still do not consider the discipline of RE to be an integrated part of software engineering. Hence, no integrated RE approach was available as a guideline in the projects.

\subsection{Effort Impacts (RQ \ref{rq:economicimpacts})}
\label{sec:resultsrq4}

We now analyse the effort spent and the efficiency of used RE processes over the identified patterns. Due to confidentiality reasons, we cannot give the exact effort data. 

\begin{figure}[htp]
\begin{center}
\includegraphics[width=0.75\textwidth]{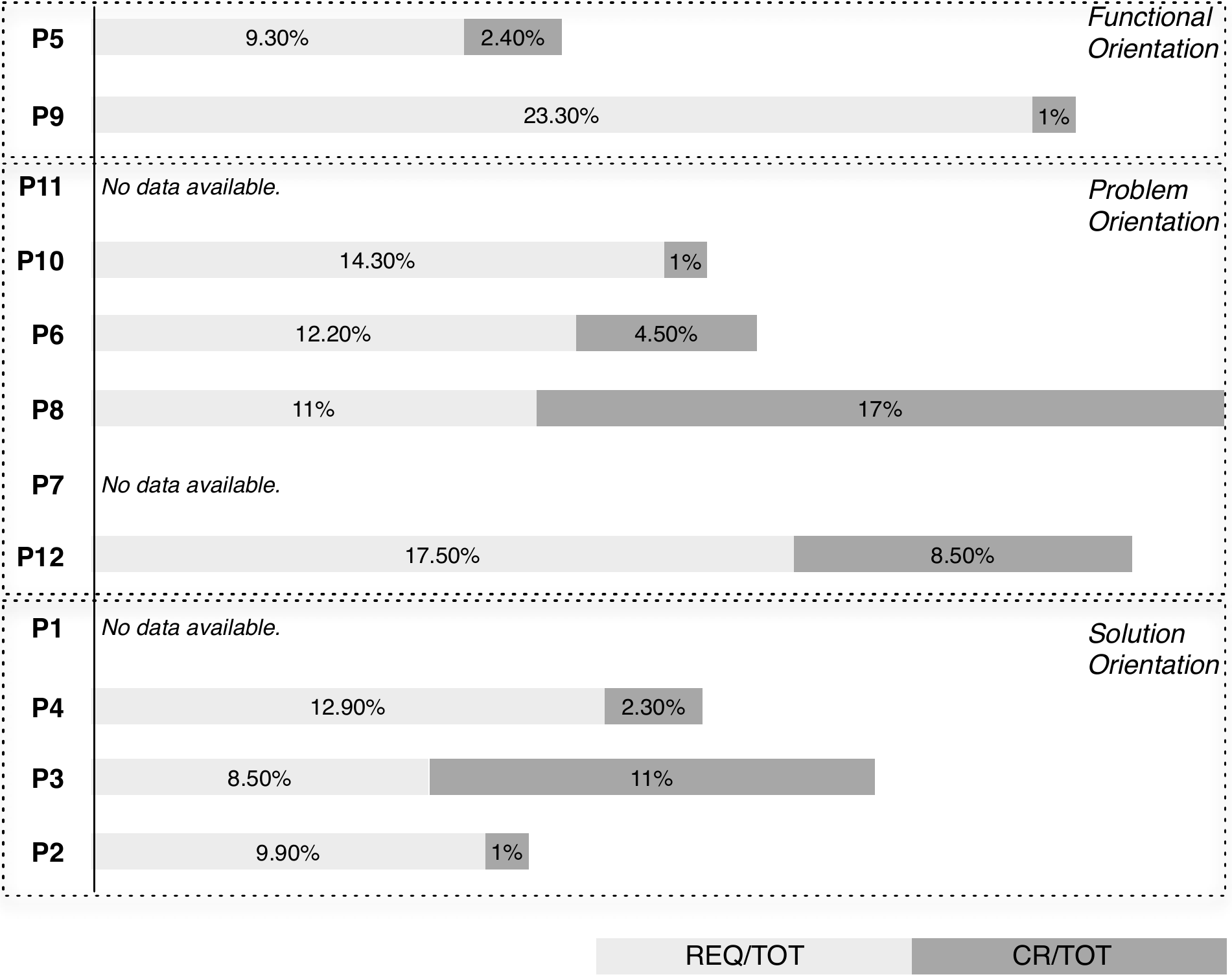}
\caption{Bar chart with the relative effort for RE (REQ/TOT) and
                change requests (CR/TOT).}
\label{fig:barchart}
\end{center}
\end{figure}
\begin{figure}[htbp]
\begin{center}
\includegraphics[width=.55\textwidth]{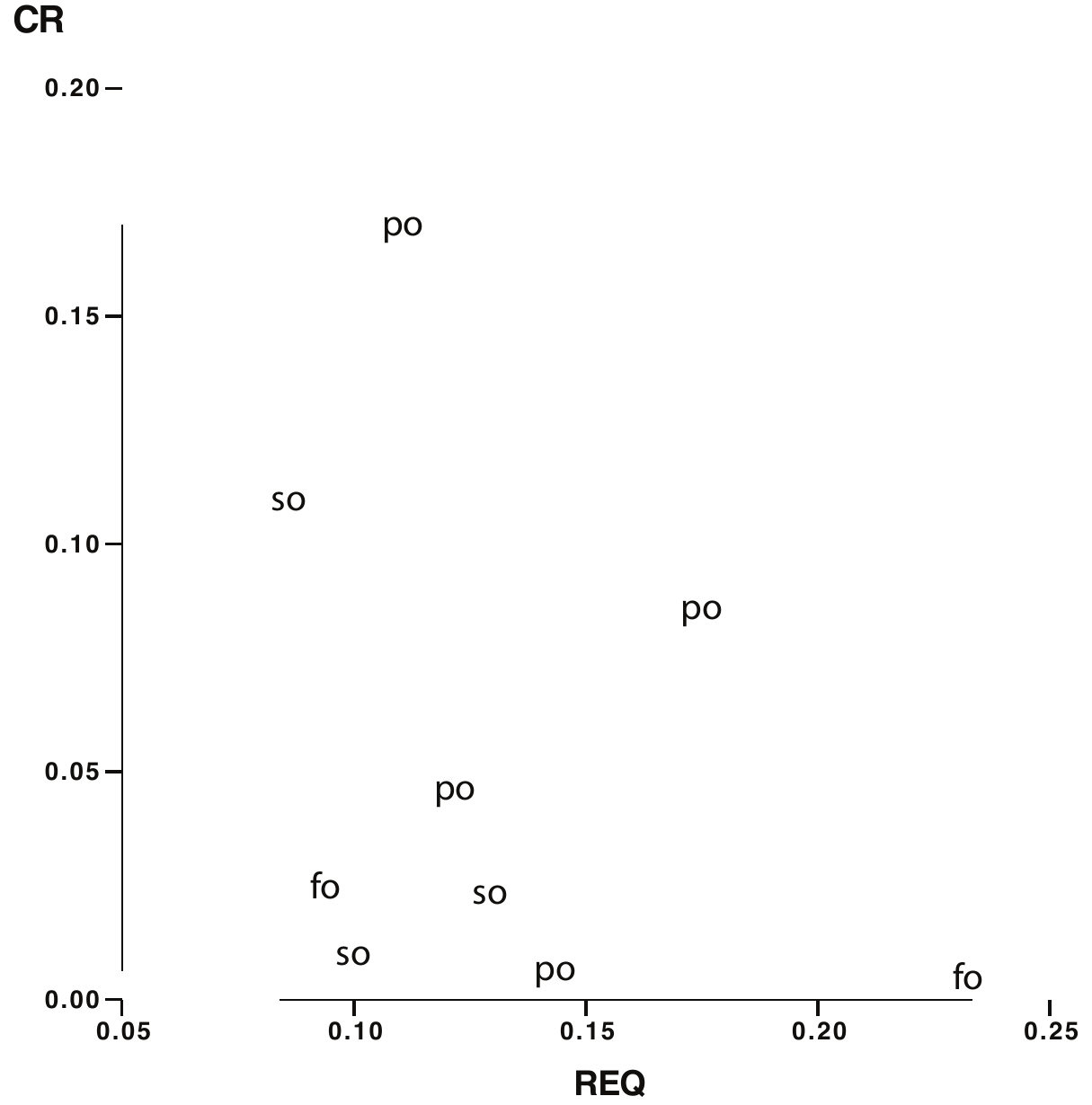}
\caption{Requirements effort REQ in relation to change request effort CR (po = problem-oriented,
                fo = functional-oriented, so = solution-oriented).}
\label{fig:req_cr}
\end{center}
\end{figure}

Figure~\ref{fig:barchart} shows the relative effort for the two categories Requirements Engineering (REQ) and change requests (CR) for each project. We use this data to analyse the relationship between these two effort categories. It is plotted in Figure~\ref{fig:req_cr} and shows a slightly negative relationship. 

The more effort goes into REQ, the less effort is needed for CR. A correlation 
analysis shows a Pearson coefficient of -0.303, but the result is not 
statistically significant at a 5\% level (p-value: 0.428). Therefore, we have an 
indication that more effort in RE can reduce the later 
effort for change requests, but we cannot confidently generalise from the data.

We have not been able to obtain the necessary data for 3 of the 12 projects. This leaves us with 2 functional-oriented projects, 4 problem-oriented projects, and 3 solution-oriented projects. Overall REQ/TOT is between 9.3\% and 23.3\%, CR/TOT is between 1\% and 17\%. In addition, Table~\ref{tab:means} shows the means of the analysed measures for each project and their totals.

\begin{table*}[htb]
\caption{Arithmetic means of the effort data for each pattern.}
\begin{center}
\begin{tabular}{lrrrr}
\hline
 & REQ/TOT & SWL/TOT & CR/TOT & CR/REQ\\
\cline{2-5}
Problem-oriented & 0.138 & 0.783 & 0.077 & 0.608\\
Functional oriented &  0.163 & 0.823 & 0.014 & 0.140\\
Solution-oriented &  0.140 & 0.848& 0.047 & 0.518\\
\hline
All & 0.132 & 0.814 & 0.053 & 0.474\\
\hline
\end{tabular}
\end{center}
\label{tab:means}
\end{table*}

To analyse the effort impacts of the patterns, we are interested in the differences between
the effort and efficiency data for the patterns. To use parameterised tests to analyse these
differences, we check whether the data is distributed normally. We perform the Shapiro-Wilk test and analyse whether the effort data is distributed normally at a significance level of 0.05.
In the test, we exclude the functional-oriented pattern, because only two projects fell into this pattern. The test shows that the effort data is not distributed normally for the chosen significance level. This means that we cannot use a t-test to compare the differences over the patterns and we have to use a nonparametric test instead. 

We use these two tests to test the null  hypotheses~\ref{hyp:req}--\ref{hyp:cr_req}, which state that there are no differences in the distribution. The alternative hypotheses are that there are differences.
The Kruskal-Wallis test is a suitable statistical test for this task.

\begin{table}[htp]
\caption{Kruskal-Wallis test on differences in the distribution of the effort data for the patterns.}
\begin{center}
\begin{tabular}{lr}
\hline
& p-value\\
\cline{2-2}
REQ/TOT &  0.425\\
CR/TOT   & 0.305 \\
SWL/TOT &  0.449 \\               
CR/REQ &  0.449 \\               
\hline
\end{tabular}
\end{center}
\label{tab:differences}
\end{table}%

Table~\ref{tab:differences} contains the p-values of the Kruskal-Wallis test for our data. 
At a significance level of 0.05, there is no support for the alternative hypotheses that there are 
significant differences between the effort and efficiency distribution of the patterns. 
All null hypotheses (\ref{hyp:req}--\ref{hyp:cr_req}) are supported by the data.
The results of the test suggest that the data comes from the same population.
Therefore, no statistically significant difference between the effort impacts
of the patterns is evident.

\subsection{Evaluation of Validity}
\label{sec:threats}
We subsequently discuss the construct validity, the internal validity and finally the external validity of the study.

\subparagraph{Construct Validity.}
The major threat to the construct validity of our study is that we analyse the
different RE execution strategies after the fact. We have not
observed the study subjects in their actual execution, but
analysed the existing artefacts and interviewed them in retrospect.
Because of this, our analysis might differ from that of a direct observation.
Analysing the existing artefacts in the content analysis, however, gives at least a basic level
of objectiveness. Furthermore, missing information was collected
directly from participating study subjects to round up the picture. In the content analysis, we also did not take into account the requirements' attributes, because they were handled differently in the projects. We see this threat, however, as a minor one, because those attributes were mainly management attributes (e.g., the priority of a requirement or the source). None of the attributes would have affected the degree of completeness in the content items given in our reference model, such as acceptance criteria or measurements for quality requirements. 

Another threat is that artefacts are not the only indicator for a specific RE execution strategy. The artefacts themselves contain no information about how they were created. To mitigate this threat, additional interviews and feedback meetings were held to add this additional information. 

For the categorisation of the projects with respect to the artefact patterns, we had to abstract from many
details of each project. Every project has its specifics because of, e.g.,
its business contexts, or the people that work in it, which all
can have a substantial influence. Some details had to
be left unconsidered for a more general analysis. We mitigated this
threat by considering the projects' details in the interviews in which we collected
project parameters. 

Similarly, regarding the effort analysis, we did not differentiate whether a customer domain was new or not. Thus, a certain learning curve in new domains (in contrast to follow-up projects) might have affected the effort. We explicitly took this aspect into account, however, as part of the project parameters and, thus, as part of the the corresponding RE execution strategies. During the effort analysis, we also had to generalise some of the effort within the SWL group that, in part, would also belong to the REQ group. For instance, effort spent for travelling was exclusively allocated to the overall software life cycle. We mitigated this threat by analysing the effort equally for all analysed projects.

\subparagraph{Internal Validity.}
In terms of the internal validity, it is possible that there were more
artefacts developed in the projects that we did not analyse but
which would have changed the classification of the project. There were,
for example, specific feature lists in some projects that were not available for analysis because of confidentiality. A similar threat comes 
from the interviews, in which the interviewees could not give us 
all their information because of confidentiality.
We mitigated this in the feedback meetings in which the study subjects
had the chance to comment on analysis and
classification. 

Furthermore, it is a threat that a large share of the information
is based on the interviews and information given by people
that participated in the projects. There is the
possibility that the information is biased. We mitigated this threat in the
preliminary interviews when we emphasised that the analysis
is not an assessment that results in statements about which
projects are good and which are bad. This lowered the barrier
to giving us complete and accurate information.

The analysis of the content items and their classification
was done subjectively by us. This holds the threat that it is not
repeatable. To mitigate that threat, we used REM as a reference model, 
classified the artefacts during the content analyses by more than one 
researcher, and discussed the results with study subjects.

Another threat is that the project parameters were gathered by discussing the influences according to the content items, and in addition, that the project parameters reflect circumstances subjectively perceived by the project participants. Hence, the project parameters and the influences documented in Table~\ref{tab:projectfactors} might be incomplete and exclusively reflect individual experiences. We did not mitigate the threat completely, since we are explicitly interested in industrial best practice and individual experiences. In addition, our interpretation of these experiences as project parameters can be subjective as well. We mitigated this threat by discussing the project parameters in the last feedback meetings with the project participants. 

Finally, further threats are at the stage of the effort analysis. The assignment of effort to the accounts by the project participants can be distorted in two ways. First, many projects were performed before standardised account names for the system were introduced in the company. We mitigated this threat by directly asking the study subjects for the corresponding account names and deviations from the actual naming convention. Second, change requests may not have been accounted for politically motivated reasons. Change requests that were not billed for whatever reason appear in the account system with negative effort. For instance, if a change request was documented with ``10 person days'', it was possible that project participants additionally accounted the same task with ``-10 person days''. We mitigated this threat by only analysing positive effort and disregarding negative ones.

\subparagraph{External Validity.}
The results of this study can only be generalised to some extent
because the major threat is that we only analysed a single company.
The results may depend to a large extent on company-specific parameters such as the development project or the corporate culture. Moreover, all analysed projects built a certain kind of business
information system. Hence, it is not clear to what extent the results
can be transferred to other categories of systems. These threats are mitigated by analysing projects from different industrial sectors and by including the company-specific development process model into the investigation. Furthermore, the analysed company is large enough to ensure a relatively low overlap between the analysed projects.

\section{Conclusions and Future Work}
\label{sec:conclusions}

In the following, we summarise our conclusions and the relation to 
existing evidence. After discussing the implications of our contribution,
we sum up its limitations, before concluding with an outline of future work.

\subsection{Summary of Conclusions}

We analysed 12 industrial projects regarding their Requirements Engineering
execution strategy. To this end, we investigated the existence and
completeness of RE artefacts referring to a generic artefact model, as well as using
interviews to gather further information. The result is an analysis of the
produced artefacts, project parameters influencing the specification of these artefacts, and a
categorisation of the projects into 3 main patterns. We found that half of the projects
act in a problem-oriented way, a third act in a functional-oriented way and a sixth in a solution-oriented way. 
We evaluated each pattern to discover differences in their efficiency and 
discovered no statistically significant difference. One may conclude that each pattern seems 
to be an appropriate response to project-specific parameters with no 
observable disadvantage.

\subsection{Relation to Existing Evidence}

In contrast to earlier studies (see Section~\ref{sec:related}), we analysed projects that were successful in terms of having released and deployed software systems to the customer that are used in production. A finding was that only 50\% of these successful projects acted in RE in a problem-oriented way. First, this manifests the idea that there still is a tendency to act in a solution-oriented way as it was already observed during the late 1970's~\cite{boehm06}. Second, we also showed that many project parameters that are considered to be the main reason for project failures, arise from the customer domain and often cannot be avoided. We found, however, that these parameters can be effectively handled as reflected in specific artefact patterns. 

The analysis was performed in a process-neutral manner investigating which RE artefacts had been specified and which influencing project parameters were affecting the specification of each single artefact. Hence, we get a more detailed view of the parameters that builds on existing ones. For instance, the \emph{Chaos Report} states in~\cite{chaos1995} that 12.4\% of project failures are caused by missing user involvement. Hull et al.~\cite{Hall.2002} discovered further organisational problems, such as inappropriate skills of project participants. Our results relate to those facts in two ways. 

First, we detailed the parameters with additional ones, like the availability of stakeholders, their technical ability, confidentiality issues or the general relationship with the customers; each having different impacts on different artefacts. Further discovered parameters that are not in the scope of current studies are, for example, the degree of user interaction, the degree of innovation (of the system) or the involvement of external suppliers. Second, we showed that parameters like these affect the creation of chosen artefacts and thereby the execution strategy, but they do not necessarily lead to a project's failure. Missing user involvement, for example, negatively affects the specification of certain content items and relates to the solution-oriented execution strategy. We achieved a more profound analysis of the produced artefacts and the underlying project parameters than provided in the study of Kamata~et~al.~\cite{IT07}. They exclusively investigated the relation of selected parts in the IEEE software requirements specification Std. 830-1998 to general project failures (with a particular focus on defects like budget and time overrun). Also, our data does not support  their observation of a balance in the depth of specifications that are produced in the projects (see Table~\ref{tab:artefacts} on page~\pageref{tab:artefacts}).

In addition to the investigated project parameters and their impacts on the artefacts created as part of specific patterns, we found that there is no significant difference between the resulting RE execution strategies regarding the spent effort or the efficiency in RE. In contrast to studies like the aforementioned ones, we discovered more project parameters and how these can be successfully handled, since the discovered artefact patterns arise as a direct ``successful'' response to these parameters with no significant losses regarding their efficiency (and, thus, project failures). 

The study by Damian~et~al.~\cite{DC06} on the influences of improvements in the RE process on further development tasks also came to a different conclusion than we did. They analysed via interviews and document inspections the payoffs of a process improvement in the RE phase within one project. After improving the RE process in that particular project, they discovered that the study participants being responsible, e.g., for project management tasks, perceived fewer field defects. Although we did not explicitly take into account the direct relationship between the RE efficiency and further development tasks, and we also abstracted from variations in the RE process definitions, we still discovered no significant difference in change requests, depending on the completeness of generated RE artefacts. 

Based on our data, we thus cannot uphold our initial assumption that front loading effort results in payoffs in later stages as motivated by studies like the one by Damian et al.~\cite{DC06}.  The objectives, the study design, and the methods used in our study, however, also differ from their study. Since they analysed one project in detail gathering a broad spectrum of experiences made by the study participants, they could focus on soft facts (e.g., team communication) and reason about the different impacts of RE on further development activities. In contrast, we analysed a broader range of development projects allowing for generalisations as part of patterns. We covered the ``soft facts'' of the projects as part of our project parameters and our effort analysis relied not on interviews, but on detailed data comprehensibly persisted in an effort accounting system that we could objectively analyse in isolation. Therefore, while Damian et al. were able to discuss the different dependencies between improvements in the RE phase and further development activities, we could objectively show how successful RE strategies are related to the chosen artefacts, rather than to the actually performed processes and used methods.

\subsection{Impact/Implications}

What can be considered a successful artefact pattern in Requirements Engineering depends on the parameters that influence the project. We found in the study that in successful projects these parameters, such as the technical knowledge of stakeholders, lead to differences in artefacts and finally artefact patterns. Each resulting execution strategy arises as a direct response to a set of project parameters, whereas we showed that the strategies have no significant differences regarding the efficiency in RE. 

Since we discovered a detailed set of project parameters, corresponding artefact patterns, and execution strategies that each can be taken as an appropriate specific-purpose reaction to individual project situations, we lay the foundation for the future elaboration of tailoring approaches. 

In such an approach, we can take the discovered project parameters and their positive or negative relationship to single artefacts as a guideline to
\begin{compactenum}
\item systematically reflect chosen project characteristics and
\item appropriately react by creating the necessary artefacts and, thus, deciding on a suitable RE execution strategy
\end{compactenum}

From a practitioner's perspective, our results can be directly taken as a framework to guide in the creation of particular content items in response to project parameters which are known bear a risk of project failure. From a researcher's perspective, we can complement available approaches in decision (support) systems for RE with the project parameters we discovered. In addition, we can complement the activity-centric area of customisation and situational method engineering -- emphasising the selection of methods according to project parameters -- with a notion of syntactic quality in the created results (see also Section~\ref{sec:future}). 

\subsection{Limitations}
The field study has only been performed at one company. If we 
expand it to other companies, this could affect the results and 
consequently the criteria for deriving the artefact patterns. We doubt that the 
basic findings would change substantially, but we would be able to make 
more detailed statements about the artefact patterns based on elaborated trends. 

In addition, all projects have been a success in the sense that they resulted in systems that are now in production on the customers' side. The different RE execution strategies are not the determinant of project success, but reflect principle ways to tackle project-specific problems. 

\subsection{Future Work}
\label{sec:future}

The importance of customisation and decision-making in Requirements Engineering has been acknowledged. There are already valuable approaches for supporting these tasks. Aurum~and~Wohlin~\cite{Aurum.2003} mapped out decision-making approaches to the RE process to understand decision-making patterns in RE activities \cite{RPAW+01}. In contrast to general activity-based customisation approaches~\cite{CCRGJ95, BABOK09, JE03, AW05},  we envision an artefact-based approach that is able to make use of the execution strategies to customise the effort according to individual project-specific parameters. 

We have proposed such an approach for the application domain of business information systems~\cite{TUM-I0929}. In this approach, we couple project parameters, as found in this field study, to selected artefacts so as to guide the systematic reflection on project characteristics, and the decision about an appropriate RE execution, respecting the necessary and possible degree of completeness in the artefacts. In~\cite{MLPW11}, we investigated the application of this approach in a case study and showed an overall improvement in the RE process with respect to the RE process previously used in the same industrial context. 

While these are only first steps, they strengthen our confidence in the suitability of the identified patterns and parameters. A further necessary step, however, consists of the extension of the found project parameters with the purpose of establishing of a comprehensive project repository, facilitating the application in real-life projects. 
Furthermore, our work can also be the foundation for developing new software requirements (process) patterns~\cite{withall07}. The focus of this study was not to bring the results in a form that they can be directly used as requirements patterns. For that, we need to document them in an appropriate template and evaluate them in different contexts.

\subsection*{Acknowledgement}

We acknowledge the effort of all the employees of Capgemini Technology Service Deutschland who participated in the study. We also want to thank G. Koller, P. Braun, J. Owen, and G. Kalus for fruitful discussions and helpful remarks on previous versions of this paper.

\bibliographystyle{model1-num-names}
\bibliography{IST}

\end{document}